\newcommand{\CLASSINPUTinnersidemargin}{0.9in}
\newcommand{\CLASSINPUToutersidemargin}{0.9in}
\documentclass[12pt, draftcls, onecolumn]{IEEEtran}
\usepackage{cite}

\ifCLASSINFOpdf
\else
  \usepackage[dvips]{graphicx}
\fi
\usepackage[cmex10]{amsmath}
\usepackage{array}
\usepackage[tight,footnotesize]{subfigure}
\usepackage{url}
\usepackage{amsthm}
\usepackage{amssymb}
\usepackage[nomarkers,nofiglist,notablist]{endfloat}
\theoremstyle{plain}
\newtheorem{lem}{\textbf{Lemma}}
\newtheorem{theorem}{Theorem}
\newtheorem{definition}{Definition}

\theoremstyle{remark}

\begin{document}
%
\title{Analysis of LT Codes with Unequal Recovery Time}

\author{
\IEEEauthorblockN{
Jesper H. S\o rensen\IEEEauthorrefmark{1},
Petar Popovski\IEEEauthorrefmark{1},
Jan \O stergaard\IEEEauthorrefmark{1},}
\IEEEauthorblockA{\IEEEauthorrefmark{1}Aalborg University, Department of Electronic Systems, E-mail: \{jhs, petarp, jo\}@es.aau.dk}
}


\maketitle

\begin{abstract}
In this paper we analyze a specific class of rateless codes, called LT codes with unequal recovery time. These codes provide the option of prioritizing different segments of the transmitted data over other. The result is that segments are decoded in stages during the rateless transmission, where higher prioritized segments are decoded at lower overhead. Our analysis focuses on quantifying the expected amount of received symbols, which are redundant already upon arrival, i.e. all input symbols contained in the received symbols have already been decoded. This analysis gives novel insights into the probabilistic mechanisms of LT codes with unequal recovery time, which has not yet been available in the literature. We show that while these rateless codes successfully provide the unequal recovery time, they do so at a significant price in terms of redundancy in the lower prioritized segments. We propose and analyze a modification where a single intermediate feedback is transmitted, when the first segment is decoded in a code with two segments. Our analysis shows that this modification provides a dramatic improvement on the decoding performance of the lower prioritized segment.
\end{abstract}
\IEEEpeerreviewmaketitle

\section{Introduction}
Rateless codes are capacity achieving erasure correcting codes. Common for all rateless codes is the ability to generate a potentially infinite amount of encoded symbols from $k$ input symbols. Decoding is possible when $(1+\epsilon)k$ encoded symbols have been received, where $\epsilon$ is close to zero. Rateless codes are attractive due to their flexibility. Regardless of the channel conditions, a rateless code will approach the channel capacity, without the need for feedback. Successful examples are LT codes \cite{fc2} and Raptor codes \cite{raptor}.

Standard rateless codes treat all data as equally important. In some applications, e.g. video streaming \cite{blostein,nybom}, this is not desirable due to dependencies between data segments. Several works address the problem of designing rateless codes which provide unequal error protection (UEP), where different data segments have different error probabilities when a certain amount of symbols have been collected. Equivalently they can also provide unequal recovery time (URT), which refers to the different amounts of symbols it requires for the different data segments to achieve the same error probability. Variants based on LT codes are found in \cite{karande,bogino,dejan,neto,rahnavard}, while \cite{cataldi} is an example using Raptor codes. Common for all approaches is the idea of biasing the random sampling in the rateless code towards the more important data. The different works are distinguished by how this is achieved. A very simple yet elegant solution, which stays true to the original LT encoding structure is found in \cite{rahnavard}. This solution replaces the originally uniform sampling of input symbols with a nonuniform one, where more important symbols are sampled with higher probability than less important symbols. The authors provide asymptotic analysis of the proposed codes for belief propagation (BP) decoding and finite-length analysis for maximum-likelihood (ML) decoding. In both cases, the UEP and URT properties are shown to be provided.

In this work, finite-length analysis is presented for BP decoding of the UEP/URT LT code from \cite{rahnavard}. We focus on the case of full recovery of the individual data segments, i.e. fixed error rate of zero. Thus, these codes are referred to as URT-LT codes. The purpose of this analysis is to quantify the amount of redundancy realizations of these codes introduce, when recovery of all data segments is desired. The analytical approach is based on a decoder state recursion, similar in structure to the one presented in \cite{karp}, though fundamentally different in its elements. To the best of our knowledge, this is the first finite-length analysis for BP decoding of this type of URT-LT codes.

Through evaluation of our analytical results, we show that the improved decoding performance of the more important data is only achieved by significantly degrading the decoding performance of less important data. Motivated by this result, we propose and analyze a modification of these codes, where successful decoding of a data segment is reported to the transmitter through a feedback channel. Adding such an intermediate acknowledgment of data is shown to have great potential in URT-LT codes. The proposal to acknowledge only a data segment is deceptively simple, however, as it can be seen from this text, the analysis is very involved. Most importantly, adding such an intermediate acknowledgment is shown to have a great potential in URT-LT codes. The assumption of having additional feedback available during a transmission may be considered a strong assumption in many communication systems. Yet it should be noted that rateless codes require a single feedback message at the end of the successful decoding. Our analysis shows that it is very beneficial to add one more such feedback message. On the other hand, often feedback is inherently available from the layers that run below the rateless code, such as e.g. the link-layer in cellular systems.


The remainder of this paper is structured as follows. Section \ref{sec:background} gives an introduction to LT codes, explaining the encoding and decoding processes and the relevant terms. In section \ref{sec:def} the notation and definitions are introduced. The analytical work is described in section \ref{sec:analysis}, followed by a numerical evaluation of the analytical results in section \ref{sec:results}. Conclusions are drawn in section \ref{sec:conclusion} and proofs of theorems and lemmas are provided in the Appendix.


 
\section{Background} \label{sec:background}
In this section an overview of standard LT codes is given, followed by a description of how to achieve URT in these codes. Assume we wish to transmit a given amount of data, which is divided into $k$ \textit{input symbols}. An encoded symbol, also called an \textit{output symbol}, is generated as the bitwise XOR of $i$ input symbols, where $i$ is found by sampling the degree distribution, $\pi(i)$. The value $i$ is referred to as the \textit{degree} of the output symbol, and all input symbols contained in an output symbol are called \textit{neighbors} of the output symbol. The degree distribution is a key element in the design of good LT codes. The encoding process of an LT code can be broken down into three steps:

\textbf{[Encoder]}
\begin{enumerate}
 \item Randomly choose a degree $i$ by sampling $\pi(i)$.
 \item Choose uniformly at random $i$ of the $k$ input symbols.
 \item Perform bitwise XOR of the $i$ chosen input symbols. The
       resulting symbol is the output symbol.
\end{enumerate}

This process can be iterated as many times as needed, which results in a rateless code.

A widely used decoder for LT codes is the belief propagation (BP) decoder. The strength of this decoder is its very low complexity \cite{raptor}. It is based on performing the reverse XOR operations from the encoding process. Assume a number of symbols have been collected and stored in a buffer, which we refer to as the \textit{cloud}. Then initially, all degree-$1$ output symbols are identified, which makes it possible to recover their corresponding neighboring input symbols. These are moved to a storage referred to as the \textit{ripple}. Symbols in the ripple are \textit{processed} one by one, which means they are XOR'ed with all output symbols, who have them as neighbors. Once a symbol has been processed, it is removed from the ripple and considered decoded. The processing of symbols in the ripple will potentially reduce some of the symbols in the cloud to degree one, in which case the neighboring input symbols are recovered and moved to the ripple. This is called a symbol \textit{release}. This makes it possible for the decoder to process symbols continuously in an iterative fashion. When a symbol is released, there is a risk that it is already available in the ripple, which means it is redundant. This part of the total redundancy is denoted $\epsilon_R$. The iterative decoding process can be explained in two steps:

\textbf{[Decoder]}
\begin{enumerate}
 \item Identify all degree-1 symbols and add the neighboring input symbols to the ripple.
 \item Process a symbol from the ripple and remove it afterwards. Go to step $1$.
\end{enumerate}

Decoding is successful when all input symbols have been recovered. If at any point before this, the ripple size equals zero, decoding has failed. If this happens, the receiver can either signal a decoding failure to the transmitter, or wait for more output symbols. In the latter case, new incoming output symbols are initially stripped for already recovered input symbols at the receiver, leaving the output symbol with what we refer to as a \textit{reduced degree}. For symbols having a reduced degree, $i$ is referred to as the \textit{original degree}. If the reduced degree is one, the symbol is added to the ripple and the iterative decoding process is restarted. In case the reduced degree is greater than one, the symbol is added to the cloud, while a reduced degree equal to zero means that the symbol is redundant. This part of the total redundancy, $\epsilon$, is denoted $\epsilon_0$, and we have that $\epsilon=\epsilon_0+\epsilon_R$.

\subsection{LT Codes with URT} \label{sec:multilayer}
In this work we will use the approach to URT proposed in \cite{uepfc1}. In this approach the uniform distribution used for selection of input symbols is replaced by a distribution which favors more important symbols. Hence, in step $2$ of the encoder, a non-uniform random selection of symbols is performed instead. This solution to URT has no impact on the decoder. We refer to these codes as URT-LT codes.

\section{Definitions and Notation}\label{sec:def}
Vectors will be denoted in bold and indexed with subscripts, e.g. $X_i$ is the $i$'th element of the vector $\pmb{X}$. The sum of all elements is denoted $\hat{X}$ and the zero vector is denoted $\pmb{0}$. Random variables are denoted with upper case letters and any realization in lower case. The probability mass function (pmf) of a random variable $X$ is denoted $f_X(x)$. For ease of notation, we will denote the conditional distribution of $X$ given $Y$ as $f_X(x|y)$ as an equivalent of $f_X(x|Y=y)$.

In URT-LT codes, the $k$ input symbols are divided into $N$ subsets $s_1$, $s_2$,..., $s_N$, each with size $\alpha_1 k$, $\alpha_2 k$,..., $\alpha_N k$, where $\sum_{j=1}^N \alpha_j = 1$. We refer to these subsets as \textit{layers}. We define the vector $\pmb{\alpha}=\left[\alpha_1, \alpha_2,...,\alpha_N\right]$. The probability of selecting input symbols from $s_j$ is $p_j(k)\alpha_j k$, such that $\sum_{j=1}^N p_j(k)\alpha_j k = 1$ and without loss of generality we assume that $p_i(k) \ge p_j(k)$ if $i < j$. Note that if $p_j(k)=\frac{1}{k}$ $\forall j$, then all data is treated equally, as in the standard single layer LT code. We define a vector, $\pmb{\beta}$, where $\beta_i=\frac{p_i(k)}{p_N(k)}$. 

An encoded symbol of a URT-LT code can be seen as having at most $N$ dimensions. The $N$-dimensional original degree is denoted $\pmb{j}$, where $j_n$ denotes the number of neighbors belonging to the $n$'th layer. Correspondingly, we refer to $\pmb{i'}$ as the reduced degree, where $i'_n$ denotes the reduced number of neighbors belonging to the $n$'th layer. Moreover, we define $\pmb{L}$, where $L_n$ denotes the number of unprocessed input symbols from the $n$'th layer. Similarly, we define $\pmb{R}$, where $R_n$ is the number of symbols in the ripple belonging to the $n$'th layer. The cloud content is denoted $\pmb{C}$, where $C_i$ is the number of symbols in the cloud having original degree $i$. Note that no differentiation between layers is made for the cloud. For the purpose of the analysis, this differentiation is not necessary in the definition of the cloud content. Instead, the differentiation will be made in the analysis. By $\mathcal{J}_i$, we denote the set of $\pmb{j}$ which satisfy $j_n \ge i'_n$, $n=1,2,...,N$, and $\hat{j}=i$.

\begin{definition}(Decoder State)
A decoder state, $\pmb{D}$, is defined by three parameters; the remaining unprocessed symbols, $\pmb{L}=\left[L_1\text{ }L_2 ... L_n\right]$, the ripple content, $\pmb{R}=\left[R_1\text{ }R_2 ... R_n\right]$, and the cloud content, $\pmb{C}=\left[C_2\text{ }C_3 ... C_k\right]$. Hence,
\begin{align}
\pmb{D} = [\pmb{L}\text{ }\pmb{R}\text{ }\pmb{C}]. \notag
\end{align}
\end{definition}

The receiver collects a number of encoded symbols, denoted $\Delta$, prior to decoding. We define the vector $\pmb{\Omega}=\left[\Omega_1,\dots,\Omega_i,\dots,\Omega_k\right]$, where $\Omega_i$ denotes the number of symbols with original degree $i$ among the $\Delta$ collected symbols. After having identified all $\Omega_1$ degree-$1$ symbols and created the initial ripple, we have what we refer to as an \textit{initial state}, whose distribution function is defined in Definition \ref{init}.

\begin{definition}(Initial State)\label{init}
An initial state, $\pmb{D}^I=[\pmb{L}^I\text{ }\pmb{R}^I\text{ }\pmb{C}^I]$, is defined as the state of the decoder after having identified the initial ripple, but before processing the first symbol. Its probability distribution function is denoted $f_{\pmb{D}^I}\left(\pmb{d}^I|\Delta\right)$ and is supported by the state space $\pmb{D}$. By $\mathcal{I}$, we denote the set of all $\pmb{d}^I$ for which $f_{\pmb{D}^I}\left(\pmb{d}^I|\Delta\right)>0$.
\end{definition}

\textit{Example:} Consider the case of $k=10$, $N=2$, $\alpha_1 k=6$ and $\alpha_2 k=4$. Decoding is attempted at $\Delta=10$ and the received output symbols have the following degrees respectively: $2,3,2,4,7,1,2,1,4,1$. Hence, $\Omega=\left[3,3,1,2,0,0,1,0,0,0\right]$. The three degree-$1$ symbols constitute the initial ripple and two of them belong to layer $1$. In this case the initial state will be as follows:

\begin{align}\label{init_example}
\pmb{L}^I&=\left[6\text{ }4\right], \notag \\
\pmb{R}^I&=\left[2\text{ }1\right], \notag \\
\pmb{C}^I&=\left[3\text{ }1\text{ }2\text{ }0\text{ }0\text{ }1\text{ }0\text{ }0\text{ }0\right].
\end{align}

With an initial state as a starting point, the decoding process can be performed. Whenever a symbol from the ripple is processed, $\hat{L}$ will decrease by one. We refer to this as a decoding step. A new decoding step can only be performed if the ripple size is greater than zero. The state distribution after $k-\hat{L}$ decoding steps, given that $\Delta$ output symbols have been collected prior to decoding, is denoted $f_{\pmb{D}^{\hat{L}}}\left(\pmb{d}^{\hat{L}}|\Delta\right)$. It is implicitly understood that the $\Delta$ output symbols, through sampling of the degree distribution, give rise to the initial state distribution, $f_{\pmb{D}^I}\left(\pmb{d}^I|\Delta\right)$, which is the starting point of the decoding. A recursive expression of $f_{\pmb{D}^{\hat{L}}}\left(\pmb{d}^{\hat{L}}|\Delta\right)$ is presented in equation \eqref{dsr}, where $\hat{L}$ is the recursion parameter. The joint state distribution is in \eqref{dsr} expressed as a function of the individual conditional distribution functions. Here $\pmb{D}^{\hat{L}}$ denotes the decoder state when $\hat{L}$ symbols remain unprocessed. Through the definition of the decoder state, we similarly have $\pmb{L}^{\hat{L}}$, $\pmb{R}^{\hat{L}}$ and $\pmb{C}^{\hat{L}}$. Note that fixing $\hat{L}$ only fixes the sum of $\pmb{L}$, thereby leaving $\pmb{L}^{\hat{L}}$ as a random variable.


\begin{align}\label{dsr}
f_{\pmb{D}^{\hat{L}}}\left(\pmb{d}^{\hat{L}}|\Delta\right) &= \sum_{\pmb{d}^{\hat{L}+1}:\pmb{r}^{\hat{L}+1}\neq\pmb{0}} f_{\pmb{R}^{\hat{L}}}\left(\pmb{r}^{\hat{L}}|\pmb{c}^{\hat{L}},\pmb{\ell}^{\hat{L}},\pmb{d}^{\hat{L}+1}\right) \notag \\ 
                                         & \quad {} \times f_{\pmb{C}^{\hat{L}}}\left(\pmb{c}^{\hat{L}}|\pmb{\ell}^{\hat{L}},\pmb{d}^{\hat{L}+1}\right) f_{\pmb{L}^{\hat{L}}}\left(\pmb{\ell}^{\hat{L}}|\pmb{d}^{\hat{L}+1}\right) \notag \\ 
                                         & \quad {} \times f_{\pmb{D}^{\hat{L}+1}}\left(\pmb{d}^{\hat{L}+1}|\Delta\right), \quad {} \mathrm{for} \text{ } \pmb{d}^{\hat{L}} \notin \mathcal{I}, \notag \\
f_{\pmb{D}^{\hat{L}}}\left(\pmb{d}^{\hat{L}}|\Delta\right) &= f_{\pmb{D}^I}\left(\pmb{d}^{\hat{L}}|\Delta\right),\qquad {} \qquad {} \mathrm{for} \text{ } \pmb{d}^{\hat{L}} \in \mathcal{I},
\end{align}
\noindent where $f_{\pmb{R}^{\hat{L}}}\left(\cdot|\cdot\right)$, $f_{\pmb{C}^{\hat{L}}}\left(\cdot|\cdot\right)$ and $f_{\pmb{L}^{\hat{L}}}\left(\cdot|\cdot\right)$ denote conditional distributions of $\pmb{R}^{\hat{L}}$, $\pmb{C}^{\hat{L}}$ and $\pmb{L}^{\hat{L}}$, respectively, and are derived in the analysis in section \ref{sec:analysis}.

In a decoding step, a number of output symbols may release and enable recovery of input symbols. We define the matrix $\pmb{M}^{\hat{L}}=\left[\pmb{M}_1^{\hat{L}},\dots,\pmb{M}_i^{\hat{L}},\dots,\pmb{M}_k^{\hat{L}}\right]$, with column vectors $\pmb{M}_i^{\hat{L}}=\left[M_{1i}^{\hat{L}},\dots,M_{ni}^{\hat{L}},\dots,M_{Ni}^{\hat{L}}\right]^T$, where $M_{ni}^{\hat{L}}$ denotes the number of releases in the $(k-\hat{L})$'th decoding step from symbols of original degree $i$, whose single remaining neighbor belongs to layer $n$. Similarly, we define the row vectors of matrix $\pmb{M}^{\hat{L}}$ as $\pmb{M}_n^{\hat{L}}=\left[M_{n1}^{\hat{L}},\dots,M_{ni}^{\hat{L}},\dots,M_{nk}^{\hat{L}}\right]$. Whenever ripple size equals zero, the decoding process stops. In this case we are left with what we refer to as a \textit{terminal state}. Its distribution function is defined in Definition \ref{ts}.

\begin{definition}(Terminal State Distribution)\label{ts}
A terminal state, $\pmb{D}^T=[\pmb{L}^T\text{ }\pmb{R}^T\text{ }\pmb{C}^T]$, is defined as a state in which $\pmb{R}=\pmb{0}$. Hence,
\begin{align}
f_{\pmb{D}^T}\left(\pmb{d}^T|\Delta\right) = \left\{ \begin{array}{ll} f_{\pmb{D}^{\hat{L}}}\left(\pmb{d}^T|\Delta\right),  &\mathrm{for}\text{ }\pmb{r}^T=\pmb{0}, \\ 0, & \mathrm{elsewhere}. \end{array} \right.
\end{align}
\end{definition}

If the decoder is in a terminal state and all $k$ input symbols have not yet been decoded, it means that more symbols must be collected in order to further progress the decoding. Once a symbol of reduced degree $1$ is received, the decoding can be restarted. The number of symbols collected while being in the terminal state $\pmb{d}^T$ is denoted $\Delta_{\pmb{d}^T}$.

\begin{figure}[t]
\centering
\includegraphics[width=0.95\columnwidth]{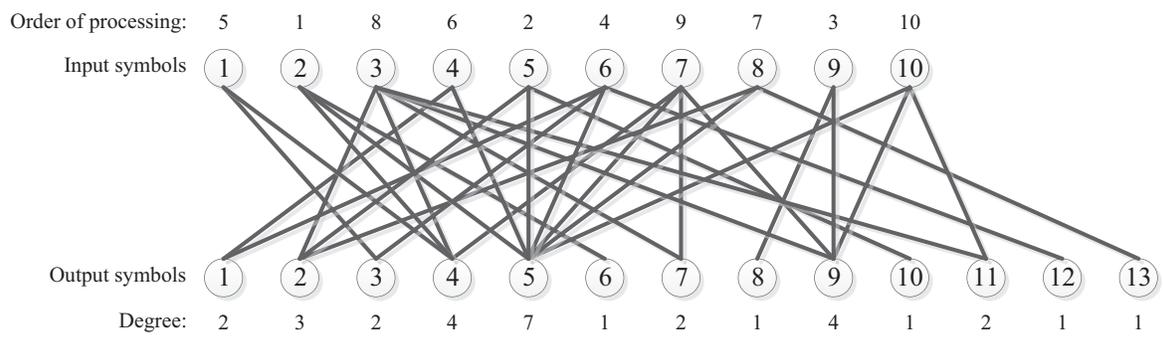}
\caption{Graph representation of the example code.}
\label{fig:graph}
\end{figure}

\begin{figure}[t]
\centering
\includegraphics[width=0.95\columnwidth]{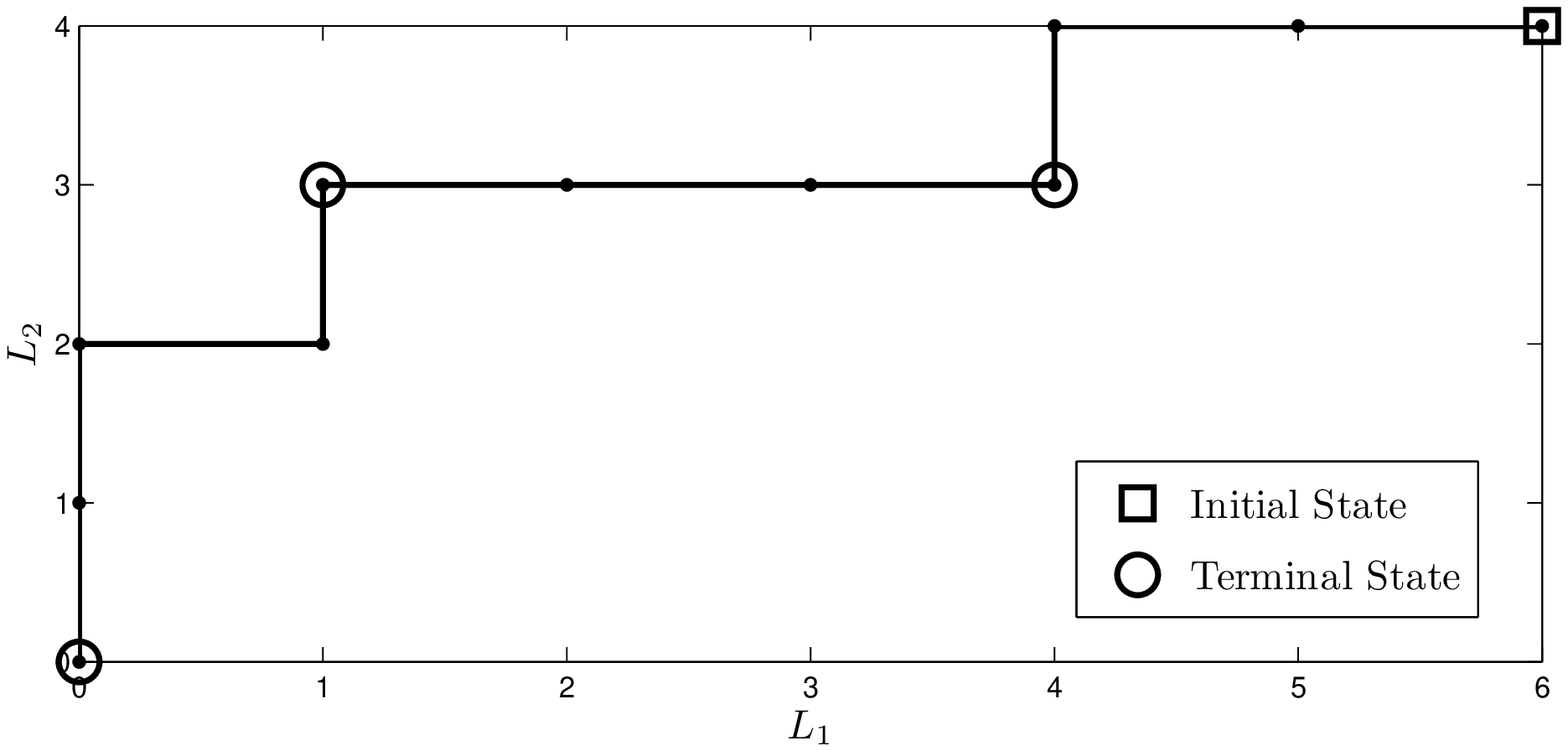}
\caption{An example of a decoder state evolution with two terminal states before successful decoding.}
\label{fig:state_ex}
\end{figure}

Table \ref{tab:state_ex} shows an example of a decoder state evolution, where each row refers to a decoder state, in which $k-\hat{L}$ symbols have been decoded and processed. The initial state is the example from \eqref{init_example}. The first decoding attempt results in a terminal state after processing three symbols. In this terminal state, two new symbols of reduced degrees $2$ and $1$ respectively are received, which enables further progress until a new terminal state is reached. Here a single new symbol of reduced degree $1$ is collected, which enables decoding of the final symbols. The individual decoding attempts are indicated with double line separations, and terminal states are highlighted in bold. Note that a decoding attempt can be interpreted as a realization of the recursion in \eqref{dsr}. The evolution of $\pmb{L}$ is illustrated in Fig. \ref{fig:state_ex} and the graph representation of this example is illustrated in Fig. \ref{fig:graph}.

We have allowed ourselves an abuse of notation, since e.g. the $T$ in $f_{\pmb{D}^T}\left(\pmb{d}^T|\Delta\right)$ does not refer to a specific value of $\hat{L}$. It is used as an indication of a certain type of state, in this case a terminal state. Hence, whenever $\pmb{D}^{X}$ is used, where $X \ne \hat{L}$, this refers to a subset of the state space, for which certain criteria are defined.

\begin{table}[ht]
\centering
\caption{Example of a decoder state evolution.}
\label{tab:state_ex}
\setlength{\tabcolsep}{2.88pt}
\begin{tabular}{|c|c|c|c|c|c|c|c|c|c|c|c|c|c|}
  \hline                       
  $k-\hat{L}$ & $L_1$ & $L_2$ & $R_1$ & $R_2$ & $C_2$ & $C_3$ & $C_4$ & $C_5$ & $C_6$ & $C_7$ & $C_8$ & $C_9$ & $C_{10}$ \\ \hline \hline
      0       &   6   &   4   &   2   &   1   &   3   &   1   &   2   &   0   &   0   &   1   &   0   &   0   &   0      \\ \hline
      1       &   5   &   4   &   1   &   1   &   3   &   1   &   2   &   0   &   0   &   1   &   0   &   0   &   0      \\ \hline
      2       &   4   &   4   &   0   &   1   &   3   &   1   &   2   &   0   &   0   &   1   &   0   &   0   &   0      \\ \hline
\textbf{3}&\textbf{4}&\textbf{3}&\textbf{0}&\textbf{0}&\textbf{3}&\textbf{1}&\textbf{2}&\textbf{0}&\textbf{0}&\textbf{1}&\textbf{0}&\textbf{0}&\textbf{0}      \\ \hline \hline
      3       &   4   &   3   &   1   &   0   &   4   &   1   &   2   &   0   &   0   &   1   &   0   &   0   &   0      \\ \hline
      4       &   3   &   3   &   2   &   0   &   2   &   1   &   2   &   0   &   0   &   1   &   0   &   0   &   0      \\ \hline
      5       &   2   &   3   &   1   &   0   &   2   &   1   &   2   &   0   &   0   &   1   &   0   &   0   &   0      \\ \hline
\textbf{6}&\textbf{1}&\textbf{3}&\textbf{0}&\textbf{0}&\textbf{2}&\textbf{1}&\textbf{2}&\textbf{0}&\textbf{0}&\textbf{1}&\textbf{0}&\textbf{0}&\textbf{0}      \\ \hline \hline
      6       &   1   &   3   &   0   &   1   &   2   &   1   &   2   &   0   &   0   &   1   &   0   &   0   &   0      \\ \hline
      7       &   1   &   2   &   1   &   0   &   2   &   0   &   2   &   0   &   0   &   1   &   0   &   0   &   0      \\ \hline
      8       &   0   &   2   &   0   &   2   &   0   &   0   &   1   &   0   &   0   &   1   &   0   &   0   &   0      \\ \hline
      9       &   0   &   1   &   0   &   1   &   0   &   0   &   0   &   0   &   0   &   0   &   0   &   0   &   0      \\ \hline
\textbf{10}&\textbf{0}&\textbf{0}&\textbf{0}&\textbf{0}&\textbf{0}&\textbf{0}&\textbf{0}&\textbf{0}&\textbf{0}&\textbf{0}&\textbf{0}&\textbf{0}&\textbf{0}     \\ \hline
\end{tabular}
\end{table}

\section{Analysis} \label{sec:analysis}
When an LT code has been partially decoded, i.e. $k'$, $0<k'<k$, input symbols have been recovered, there is a probability that the reduced degree of a new incoming symbol is zero. If this happens, the symbol is redundant and therefore discarded. Naturally this probability is of great importance in the search of well performing LT codes. In this analysis we derive the reduced degree distribution for an LT code with URT and use it to show the probability of redundancy in such codes. We also show how simple use of ACK can significantly decrease this probability.

\subsection{URT-LT Codes Without Feedback}\label{nofb}
Initially, we express in Theorem \ref{nrdd} the reduced degree distribution, $\pi_{\beta}'(\pmb{i'},\pmb{\ell})$, for a certain $\beta$-value and as a function of the number of unprocessed symbols from individual layers, $\pmb{\ell}$.

\begin{theorem}(N-Layer Reduced Degree Distribution)\label{nrdd}
Given that the encoder applies $\pi(i)$ in an N-layer URT-LT code with parameters, $\pmb{\alpha}$ and $\pmb{\beta}$, where $\ell_n$ symbols remain unprocessed from the $n$'th layer, $n=1,2...N$, the reduced degree distribution, $\pi_{\beta}'(\pmb{i'},\pmb{\ell})$, is found as
\begin{align}
\scriptstyle \pi_{\beta}'(\pmb{i'},\pmb{\ell}) \hspace{0.1cm} = \hspace{0.1cm} & \sum_{\scriptscriptstyle i=\hat{i}'}^{\scriptscriptstyle \hat{i}'+k-\hat{\ell}} \bigg(\scriptstyle\pi(i) \displaystyle\sum_{\scriptscriptstyle \pmb{j}\in\mathcal{J}_i} \bigg( \scriptstyle \Phi(\pmb{j},i,\pmb{\alpha} k,\pmb{\beta}) \displaystyle\prod_{n=1}^N \scriptstyle\frac{\binom{\ell_n}{i_n'}\binom{\alpha_n k-\ell_n}{j_n-i_n'}}{\binom{\alpha_n k}{j_n}} \bigg) \bigg)\notag \\ 
& \scriptstyle \hspace{0.2cm}\mathrm{for} \hspace{0.2cm} \ell_n < i_n' \le \alpha_n k, \hspace{0.2cm} n=1,2,...,N, \notag \\
\scriptstyle \pi_{\beta}'(\pmb{i'},\pmb{\ell}) \hspace{0.1cm} = \hspace{0.1cm} & \scriptstyle 0 \hspace{0.2cm} \mathrm{elsewhere}. \notag
\end{align}
\noindent where $\Phi$ is Wallenius' noncentral hypergeometric distribution. \qed
\end{theorem}

When evaluating $\pi_{\beta}'(\pmb{i'},\pmb{\ell})$ at $\pmb{i'}=\pmb{0}$ we get an interesting quantity. At a given terminal state, $\pmb{d}^T$, during transmission, when $\ell_n^T$ symbols remain unprocessed from the $n$'th layer, $n=1,2...N$, $\pi_{\beta}'\left(\pmb{0},\pmb{\ell}^T\right)$ is the probability that the next received symbol is redundant. This is a key element of this analysis, since it enables us to evaluate the expected value of $\epsilon_0$. This requires that we know how $\pmb{\ell}^T$ evolves as $\Delta$ increases. The derivation of this, is the goal of the further analysis.

\begin{figure}[t]
\centering
\includegraphics[width=0.95\columnwidth]{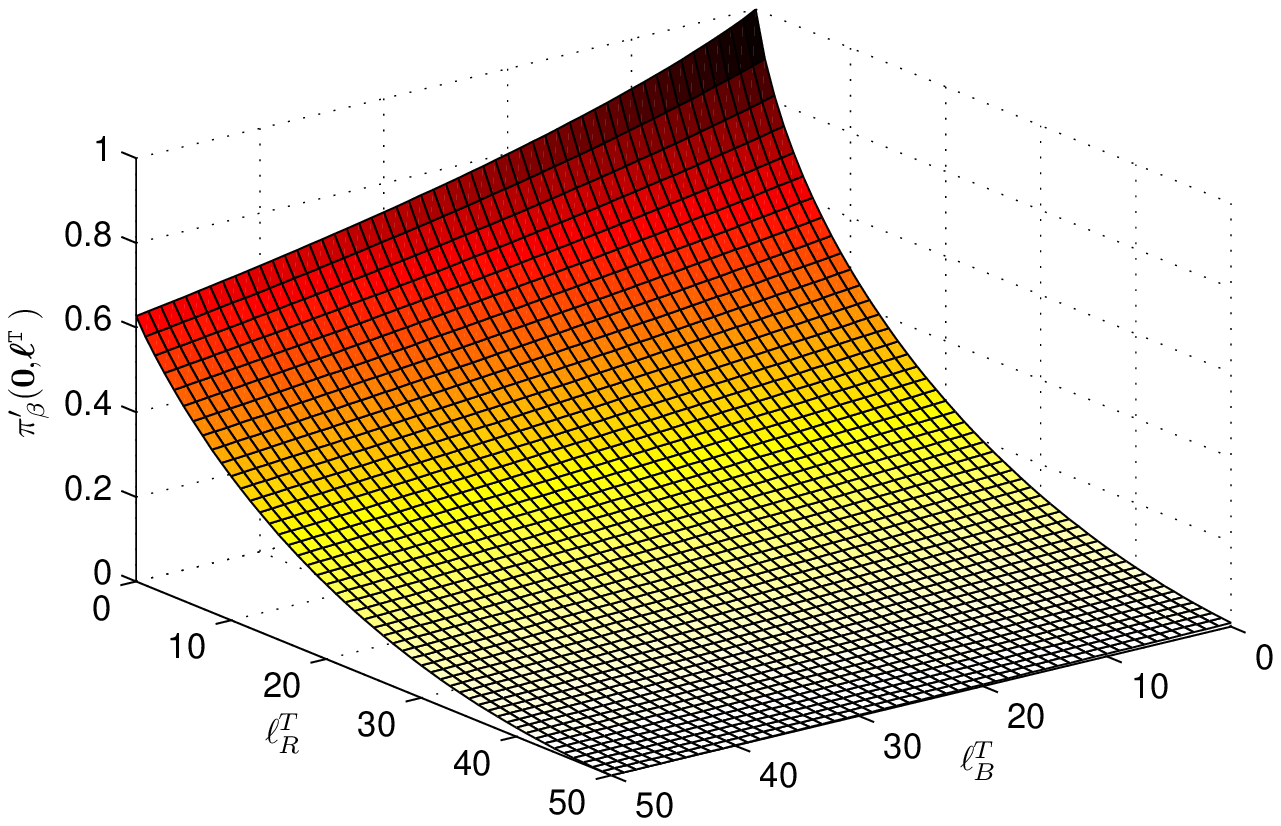}
\caption{The probability of receiving a symbol with reduced degree zero, $\pi_{\beta}'\left(\pmb{0},\pmb{\ell}\right)$, for a two-layer LT code with parameters $\alpha=0.5$, $\beta=9$ and $k=100$.}
\label{fig:twolayerred}
\end{figure}

In the rest of the analysis, we will treat the case of $N=2$, where we refer to the layers as \textit{base layer} ($n=1$) and \textit{refinement layer} ($n=2$). In Fig. \ref{fig:twolayerred}, $\pi_{\beta}'(\pmb{0},\pmb{\ell})$ has been plotted as a function of $L_B$ and $L_R$, the number of undecoded symbols from the base layer and the refinement layer, respectively. The parameters, $\alpha_B=0.5$, henceforth denoted as $\alpha$, $\alpha_R=1-\alpha$, $\beta=\frac{p_B}{p_R}=9$ and $k=100$ have been chosen. An optimized degree distribution is not provided in \cite{rahnavard} and such an optimization is out of the scope of this paper, thus we have chosen the Robust Soliton distribution (RSD). The RSD is the de facto standard degree distribution for LT codes and was originally proposed in \cite{fc2}. The plot shows that the probability of redundancy increases faster for decreasing $L_B$ than for decreasing $L_R$, which is expected since the base layer symbols are more likely to occur as neighbors.

We will now use $\pi_{\beta}'\left(\pmb{0},\pmb{\ell}\right)$ to derive the expected amount of redundancy, $\mathbb{E}\left[k \epsilon_0(\Delta_{max})\right]$, due to symbols having reduced degree zero, when a maximum of $\Delta_{max}$ symbols are collected. In order to do this, we must find the expected number of symbols received in each possible state. In this regard, we note that the $\Delta$'th symbol is received in state $\pmb{d}^T$, if the decoding of the first $\Delta-1$ symbols resulted in the terminal state $\pmb{d}^T$. Hence, the expected number of symbols, $\mathbb{E}\left[\Delta_{\pmb{d}^T}(\Delta_{max})\right]$, received while being in state $\pmb{d}^T$, equals the expected number of times decoding fails in that state. When having $\mathbb{E}\left[\Delta_{\pmb{d}^T}(\Delta_{max})\right]$, it is easy to obtain $\mathbb{E}\left[k \epsilon_0(\Delta_{max})\right]$ by multiplying with $\pi_{\beta}'\left(\pmb{0},\pmb{\ell}^T\right)$ and summing over all $\pmb{d}^T$, for which $\pmb{\ell}^T\neq\pmb{0}$. Note that $\pi_{\beta}'\left(\pmb{0},\pmb{\ell}^T\right)$ only depends on $\pmb{d}^T$ and not $\Delta$, hence it can be left outside the innermost expectation. The expected amount of redundancy is expressed in Theorem \ref{red1}.

\begin{theorem}(Redundancy in Two-Layer URT-LT Code)\label{red1}
In a two-layer URT-LT code using any degree distribution, $\pi(i)$, the expected total amount of redundancy, $\mathbb{E}\left[k \epsilon_0(\Delta_{max})\right]$, due to reduced degree zero, is:
\begin{align}
\mathbb{E}\left[k \epsilon_0(\Delta_{max})\right] &= \sum_{\pmb{d}^T:\pmb{\ell}^T\neq\pmb{0}} \mathbb{E}\left[\Delta_{\pmb{d}^T}(\Delta_{max})\right] \pi_{\beta}'\left(\pmb{0},\pmb{\ell}^T\right), \notag \\
\mathbb{E}\left[\Delta_{\pmb{d}^T}(\Delta_{max})\right] &= \sum_{\Delta=1}^{\Delta_{max}} f_{\pmb{D}^T}\left(\pmb{d}^T|\Delta\right). \notag 
\end{align} \qed
\end{theorem}

Theorem \ref{red1} makes use of the recursive state distribution function in \eqref{dsr}. This function depends on the initial state distribution, $f_{\pmb{D}^I}\left(\pmb{d}^I|\Delta\right)$, which has not yet been derived, as is the case for the conditional distributions of the three dimensions of the decoder state, $f_{\pmb{R}^{\hat{L}}}\left(\cdot|\cdot\right)$, $f_{\pmb{C}^{\hat{L}}}\left(\cdot|\cdot\right)$ and $f_{\pmb{L}^{\hat{L}}}\left(\cdot|\cdot\right)$. To support the derivation of these, we state Lemma \ref{ripadds}, which expresses the distribution, $f_{\hat{M}'_n}\left(\hat{m}_n'|\hat{m}_n,\ell_n,r_n\right)$, of $\hat{M}_n'$, the number of symbols added to the $n$'th dimension of the ripple, when the ripple size in this dimension is $r_n$, $\ell_n$ symbols remain unprocessed in the $n$'th layer and $\hat{m}_n$ symbols have been released and have a neighbor from the $n$'th layer.

\begin{lem} (Ripple Influx)\label{ripadds}
When $\hat{m}_n$ symbols have been released, i.e. have only one neighbor, which belongs to the $n$'th layer, at a point where $\ell_n$ symbols remain unprocessed from the $n$'th layer and the ripple contains $r_n$ symbols in the $n$'th dimension, the random variable $\hat{M}_n'$ denotes how many of those will be added to the ripple. It has the following distribution:
\begin{align}
f_{\hat{M}'_n}\left(\hat{m}_n'|\hat{m}_n,\ell_n,r_n\right)&=\sum_{q=\hat{m}_n'}^{min(\hat{m}_n,\ell_n)} \frac{\binom{\ell_n}{q} Z_q(\hat{m}_n)}{\ell_n^{\hat{m}_n}} \Upsilon (\hat{m}_n',\ell_n,\ell_n-r_n,q), \notag \\
Z_q(\hat{m}_n) &= \sum_{p=0}^q \binom{q}{p} (q-p)^{\hat{m}_n}(-1)^p \notag,
\end{align}
\noindent where $\Upsilon$ is the hypergeometric distribution.\qed
\end{lem}

For the initial state distribution, it is given that $\pmb{L}^I=[\alpha k,\text{ }(1-\alpha) k]$, however, $\pmb{R}^I$ and $\pmb{C}^I$ are random variables, which depend on the realization, $\pmb{\omega}$, of $\pmb{\Omega}$. These realizations follow the multinomial distribution. We have that $C^I_i=\omega_i$ for $i>1$. Symbols of degree $1$ will release immediately, hence $\hat{M}_1^{k}=\omega_1$. Each released symbol will have a base layer symbol as neighbor with probability $\frac{\beta\alpha}{\beta\alpha+\left(1-\alpha\right)}$, hence the binomial distribution describes the distinction of the $\hat{M}_1^{k}$ releases among layers, since $\hat{M}_1^{k}=M_{B1}^{k}+M_{R1}^{k}$. The releases within a layer will potentially result in the recovery of a new symbol and thereby contribute to $\pmb{R}^I$. Lemma \ref{ripadds} is thus used to express the distribution of $\pmb{R}^I$ given $M_{B1}^{k}$ and $M_{R1}^{k}$. We then get the following initial state distribution:

\begin{align}\label{init_state}
f_{\pmb{D}^I}\left(\pmb{d}^I|\Delta\right) &= \sum_{m_{B1}^{k}=0}^{\hat{m}_1^{k}} \mu\left([\hat{m}_1^{k}\text{ }\pmb{c}^I],\Delta,\pi(i)\right) \theta \left(m_{B1}^{k},\hat{m}_1^{k},\frac{\beta\alpha}{\beta\alpha+\left(1-\alpha\right)} \right) \notag \\
                         & \quad {} \times f_{\hat{M}_n'}\left(r^I_B|m_{B1}^{k},\alpha k,0\right) f_{\hat{M}_n'}\left(r^I_R|\hat{m}_1^{k}-m_{B1}^{k},(1-\alpha)k,0\right),
\end{align}
\noindent where $\theta$ is the binomial distribution, $\mu$ is the multinomial distribution and its realizations are constrained by $\Delta = \hat{m}_1^{k}+\hat{c}^I$.

Lemma \ref{nextproc} introduces $f_{\pmb{L}^{\hat{L}}}\left(\pmb{\ell}^{\hat{L}}|\pmb{d}^{\hat{L}+1}\right)$, which expresses the probability that the next processed symbol is a base layer symbol and the probability that it is a refinement layer symbol.

\begin{lem} (Next Processed Symbol)\label{nextproc}
Given a decoder state $\pmb{d}^{\hat{L}+1}$, the probability that the next processed symbol is a base layer symbol $\left(\ell^{\hat{L}}_B=\ell^{\hat{L}+1}_B-1,\text{ }\ell^{\hat{L}}_R=\ell^{\hat{L}+1}_R\right)$ and the probability that it is a refinement layer symbol $\left(\ell^{\hat{L}}_B=\ell^{\hat{L}+1}_B,\text{ }\ell^{\hat{L}}_R=\ell^{\hat{L}+1}_R-1\right)$, is:
\begin{align}
f_{\pmb{L}^{\hat{L}}}\left(\pmb{\ell}^{\hat{L}}|\pmb{d}^{\hat{L}+1}\right) = \left\{ \begin{array}{ll} \frac{r^{\hat{L}+1}_B}{r^{\hat{L}+1}_B+r^{\hat{L}+1}_R}, \text{ } \mathrm{for} \text{ } \ell^{\hat{L}}_B=\ell^{\hat{L}+1}_B-1 \text{ } \mathrm{and} \text{ } \ell^{\hat{L}}_R=\ell^{\hat{L}+1}_R, \\ \\ \frac{r^{\hat{L}+1}_R}{r^{\hat{L}+1}_B+r^{\hat{L}+1}_R}, \text{ } \mathrm{for} \text{ } \ell^{\hat{L}}_B=\ell^{\hat{L}+1}_B \text{ } \mathrm{and} \text{ } \ell^{\hat{L}}_R=\ell^{\hat{L}+1}_R-1. \end{array} \right. \notag
\end{align}\qed
\end{lem}

Next step is to find the distribution of the releases during decoding, since they determine the dynamics of the buffer content and are the basis of the ripple influx. Lemma \ref{relprob} expresses the probability that a symbol of degree $i$ is released in a certain decoding step. 

\begin{lem} (Release Probability in Two-Layer URT-LT Code)\label{relprob}
In a two-layer URT-LT code with parameters $\alpha$, $\beta$ and $k$, the prior probability, $q_{XY}$, that a symbol of degree $i$ is released as a symbol from layer $Y$, after processing a symbol from layer $X$, when $\ell_B$ and $\ell_R$ symbols remain unprocessed from base layer and refinement layer respectively, is:
\begin{align}
q_{BB}(i,\pmb{\ell}) = \sum_{j=0}^i \left( \Phi(j,i,\alpha k,\beta) \frac{\ell_B \binom{\alpha k-\ell_B-1}{j-2} \binom{(1-\alpha) k-\ell_R}{ i-j}}     {\binom{\alpha k}{j} \binom{(1-\alpha) k}{i-j}} \right) \notag \\
q_{BR}(i,\pmb{\ell}) = \sum_{j=0}^i \left( \Phi(j,i,\alpha k,\beta) \frac{\ell_R \binom{\alpha k-\ell_B-1}{j-1} \binom{(1-\alpha) k-\ell_R}{ i-j-1}}   {\binom{\alpha k}{j} \binom{(1-\alpha) k}{i-j}} \right) \notag \\
q_{RB}(i,\pmb{\ell}) = \sum_{j=0}^i \left( \Phi(j,i,\alpha k,\beta) \frac{\ell_B \binom{\alpha k-\ell_B}{j-1}   \binom{(1-\alpha) k-\ell_R-1}{ i-j-1}} {\binom{\alpha k}{j} \binom{(1-\alpha) k}{i-j}} \right) \notag \\
q_{RR}(i,\pmb{\ell}) = \sum_{j=0}^i \left( \Phi(j,i,\alpha k,\beta) \frac{\ell_R \binom{\alpha k-\ell_B}{j}     \binom{(1-\alpha) k-\ell_R-1}{ i-j-2}} {\binom{\alpha k}{j} \binom{(1-\alpha) k}{i-j}} \right) \notag
\end{align}
\noindent where $\Phi$ is Wallenius' noncentral hypergeometric distribution.\qed
\end{lem}

When $\hat{M}^{\hat{L}}_i$ symbols are released in the $(k-\hat{L})$'th decoding step, we have that $C^{\hat{L}}_i=C^{\hat{L}+1}_i-\hat{M}^{\hat{L}}_i$, which allows us to express the cloud development recursively, if we derive the distribution of $\hat{M}^{\hat{L}}_i$. This can be done using Lemma \ref{relprob} and the binomial distribution, which gives us the cloud development expressed in Lemma \ref{rels}.

\begin{lem} (Cloud Development in a Decoding Step)\label{rels}
Given a decoder state $\pmb{d}^{\hat{L}+1}$, with a cloud content $\pmb{c}^{\hat{L}+1}$, the probability of having a cloud content of $\pmb{c}^{\hat{L}}$, after processing a symbol from either the base layer $\left(\ell^{\hat{L}}_B=\ell^{\hat{L}+1}_B-1,\text{ }\ell^{\hat{L}}_R=\ell^{\hat{L}+1}_R\right)$ or the refinement layer $\left(\ell^{\hat{L}}_B=\ell^{\hat{L}+1}_B,\text{ }\ell^{\hat{L}}_R=\ell^{\hat{L}+1}_R-1\right)$, is:
\begin{align}
f_{\pmb{C}^{\hat{L}}}\left(\pmb{c}^{\hat{L}}|\pmb{\ell}^{\hat{L}},\pmb{d}^{\hat{L}+1}\right) &= \prod_i \theta\left(\hat{m}^{\hat{L}}_i,c^{\hat{L}+1}_i,\frac{q\left(i,\pmb{\ell}^{\hat{L}}\right)}{\sum_{\ell_R=0}^{\ell_R^{\hat{L}}} q\left(i,[0\text{ }\ell_R]\right) + \sum_{\ell_B=1}^{\ell_B^{\hat{L}}} q\left(i,[\ell_B\text{ }\ell_R^{\hat{L}}]\right)}\right) \notag \\
q\left(i,\pmb{\ell}^{\hat{L}}\right) &= \left\{ \begin{array}{ll} q_{BB}\left(i,\pmb{\ell}^{\hat{L}}\right)+q_{BR}\left(i,\pmb{\ell}^{\hat{L}}\right) \text{ } \mathrm{for} \text{ } \ell^{\hat{L}}_B=\ell^{\hat{L}+1}_B-1 \text{ } \mathrm{and} \text{ } \ell^{\hat{L}}_R=\ell^{\hat{L}+1}_R,\\ \\
q_{RB}\left(i,\pmb{\ell}^{\hat{L}}\right)+q_{RR}\left(i,\pmb{\ell}^{\hat{L}}\right) \text{ } \mathrm{for} \text{ } \ell^{\hat{L}}_B=\ell^{\hat{L}+1}_B \text{ } \mathrm{and} \text{ } \ell^{\hat{L}}_R=\ell^{\hat{L}+1}_R-1, \end{array} \right. \notag \\
\hat{m}^{\hat{L}}_i &= c^{\hat{L}+1}_i-c^{\hat{L}}_i, \notag
\end{align}
\noindent where $\theta$ is the binomial distribution, $q\left(i,\pmb{\ell}^{\hat{L}}\right)$ is the probability that an output symbol of degree $i$ is released when $\pmb{L}=\pmb{\ell}^{\hat{L}}$ and $q_{XY}$ are given in Lemma \ref{relprob}.\qed
\end{lem}

Lemma \ref{rels} expresses the probability distribution of the cloud and thereby the probability distribution of the symbol releases in a decoding step. However, a released symbol is not guaranteed to be added to the ripple. Some released symbols might be identical and some might already be in the ripple. Lemma \ref{ripadds} can be applied to express the distribution of the number of symbols added to the two dimensions of the ripple. For this purpose, we introduce $q_{XB}\left(i,\pmb{\ell}^{\hat{L}}\right)$, which is the probability that a symbol releases as a base layer symbol, given that it has been released. It is given by:


\begin{align}
q_{XB}\left(i,\pmb{\ell}^{\hat{L}}\right) = \left\{ \begin{array}{ll} \frac{q_{BB}(i,\pmb{\ell}^{\hat{L}})}{q_{BB}(i,\pmb{\ell}^{\hat{L}})+q_{BR}(i,\pmb{\ell}^{\hat{L}})}  \text{ } \mathrm{for} \text{ } \ell^{\hat{L}}_B=\ell^{\hat{L}+1}_B-1 \text{ } \mathrm{and} \text{ } \ell^{\hat{L}}_R=\ell^{\hat{L}+1}_R,\\ \\
\frac{q_{RB}(i,\pmb{\ell}^{\hat{L}})}{q_{RB}(i,\pmb{\ell}^{\hat{L}})+q_{RR}(i,\pmb{\ell}^{\hat{L}})} \text{ } \mathrm{for} \text{ } \ell^{\hat{L}}_B=\ell^{\hat{L}+1}_B \text{ } \mathrm{and} \text{ } \ell^{\hat{L}}_R=\ell^{\hat{L}+1}_R-1. \end{array} \right.
\end{align}

We now note that $M^{\hat{L}}_{Bi}$ follows the binomial distribution with parameters $\hat{m}^{\hat{L}}_i$ and $q_{XB}\left(i,\pmb{\ell}^{\hat{L}}\right)$, i.e. $\theta\left(m^{\hat{L}}_{Bi},\hat{m}^{\hat{L}}_i,q_{XB}\left(i,\pmb{\ell}^{\hat{L}}\right)\right)$. Moreover, we have that the distribution of $\hat{M}^{\hat{L}}_{B}$ is a convolution of these binomial distributions over all $i$. Hence,

\begin{align}\label{mb}
f_{\hat{M}^{\hat{L}}_{B}}\left(\hat{m}^{\hat{L}}_{B}|\pmb{c}^{\hat{L}},\pmb{c}^{\hat{L}+1}\right) &= \coprod_{i=1}^k \theta \left(m^{\hat{L}}_{Bi},\hat{m}^{\hat{L}}_i,q_{XB}\left(i,\pmb{\ell}^{\hat{L}}\right)\right), \notag \\
\hat{m}^{\hat{L}}_i &= c^{\hat{L}+1}_i-c^{\hat{L}}_i,
\end{align}
\noindent where $\theta$ is the binomial distribution and $\displaystyle \coprod_{i=1}^k$ denotes a $k$-way convolution of the binomial distributions with $i=1,2,...,k$.

We denote the number of base (refinement) layer symbols, which are added to the ripple, $\hat{M}^{\hat{L}'}_{B}$ $\left(\hat{M}^{\hat{L}'}_{R}\right)$, and can then express the probability distribution of the ripple size after the processing of a new symbol in Lemma \ref{ripdev}.

\begin{lem} (Ripple Development in a Decoding Step) \label{ripdev}
Given a decoder state $\pmb{d}^{\hat{L}+1}$, containing a ripple size of $\pmb{r}^{\hat{L}+1}$, the probability of having a ripple size of $\pmb{r}^{\hat{L}}$, after the next decoding step resulting in the cloud $\pmb{c}^{\hat{L}}$, is:
\begin{align}
f_{\pmb{R}^{\hat{L}}}\left(\pmb{r}^{\hat{L}}|\pmb{c}^{\hat{L}},\pmb{\ell}^{\hat{L}},\pmb{d}^{\hat{L}+1}\right) &= f_{\hat{M}'_n}\left(\hat{m}^{\hat{L}'}_{B}|\hat{m}^{\hat{L}}_{B},\ell_B^{\hat{L}},r_B^{\hat{L}+1}-\left(\ell^{\hat{L}+1}_B-\ell^{\hat{L}}_B\right)\right) \notag \\ & \quad {} \times f_{\hat{M}'_n}\left(\hat{m}^{\hat{L}'}_{R}|\hat{m}^{\hat{L}}-\hat{m}^{\hat{L}}_{B},\ell_R^{\hat{L}},r_R^{\hat{L}+1}-\left(\ell^{\hat{L}+1}_R-\ell^{\hat{L}}_R\right)\right) \notag \\ 
& \quad {} \times f_{\hat{M}^{\hat{L}}_{B}}\left(\hat{m}^{\hat{L}}_{B}|\pmb{c}^{\hat{L}},\pmb{c}^{\hat{L}+1}\right), \notag \\
r^{\hat{L}}_B &= \hat{m}^{\hat{L}'}_{B} + r^{\hat{L}+1}_B - \left(\ell^{\hat{L}+1}_B-\ell^{\hat{L}}_B\right), \notag \\
r^{\hat{L}}_R &= \hat{m}^{\hat{L}'}_{R} + r^{\hat{L}+1}_R - \left(\ell^{\hat{L}+1}_R-\ell^{\hat{L}}_R\right). \notag
\end{align}\qed
\end{lem}

Lemmas \ref{ripadds} through \ref{ripdev} provide the necessary support for Theorem \ref{red1}. The case where feedback is applied is analyzed in the following subsection.

\subsection{URT-LT Codes With Feedback}
In this subsection we treat the case where an intermediate feedback message is applied during the transmission. This message tells the transmitter that the base layer has been decoded, i.e. it works as an acknowledgment of the base layer. The transmitter adapts by excluding the base layer symbols from the random selection in step $2$ of the encoder. This means that only refinement layer symbols are included in future encoding. The feedback message is assumed to be perfect, i.e. zero error probability and delay. In the event where the refinement layer is decoded before the base layer, no intermediate feedback is transmitted.

In the case of feedback, we can divide the transmission into two phases; one before feedback (phase $1$) and one after (phase $2$). The number of symbols collected in phase $1$ is denoted $\Delta^1$ and the total number of symbols collected in both phases is denoted $\Delta^2$. Hence, the number of symbols collected in phase $2$ is $\Delta^2-\Delta^1$. One of the main differences between the two phases is the initial state distribution. In phase $1$ it is equivalent to the case without feedback, whereas in phase $2$ the initial state will be the result of adding $\Delta^2-\Delta^1$ symbols to the outcome of phase $1$ and identifying the new ripple. Moreover, in phase $2$ the encoder only considers refinement layer symbols, which is the equivalent of $\beta=0$, thus entailing the reduced degree distribution $\pi_{0}'\left(\pmb{0},\pmb{\ell}^T\right)$. Phase $1$ continues as long as $\ell_B^T\ne 0$ and phase $2$ continues as long as $\ell_R^T \ne 0$.

The expected redundancy, caused by symbols of reduced degree zero, is denoted $\mathbb{E}\left[k \epsilon_0^F(\Delta_{max}) \right]$ and is found using the same approach as in Theorem \ref{red1}. First we find the expected number of times the decoder fails in any state $\pmb{d}^T$ in both phase $1$ and phase $2$. This will provide the expected numbers, $\mathbb{E}\left[\Delta_{\pmb{d}^T}^1(\Delta_{max})\right]$ and $\mathbb{E}\left[\Delta_{\pmb{d}^T}^2(\Delta_{max})\right]$, of symbols received in any such state in the two phases. We then multiply with the probability that the next symbol is redundant, $\pi_{\beta}'\left(\pmb{0},\pmb{\ell}^T\right)$, and sum over all terminal states $\pmb{d}^T$ for which $\ell_B^T\ne 0$ in phase $1$, since this is required for phase $1$ to continue. Similarly, we multiply with $\pi_{0}'\left(\pmb{0},\pmb{\ell}^T\right)$ and sum over all $\pmb{d}^T$ for which $\ell_R^T \ne 0$ in phase $2$. The expected redundancy is formally expressed in Theorem \ref{red2}.  

\begin{theorem}(Redundancy in Two-Layer URT-LT Code with Feedback)\label{red2}
In a two-layer URT-LT code using any degree distribution, $\pi(i)$, and a single feedback message when the base layer has been decoded, the expected total amount of redundancy, $\mathbb{E}\left[k \epsilon_0^F(\Delta_{max}) \right]$, due to reduced degree zero, is:
\begin{align}
\mathbb{E}\left[k \epsilon_0^F(\Delta_{max}) \right] &= \sum_{\pmb{d}^T:\ell_B^T \ne 0} \mathbb{E}\left[\Delta_{\pmb{d}^T}^1(\Delta_{max})\right] \pi_{\beta}'\left(\pmb{0},\pmb{\ell}^T\right) + \sum_{\pmb{d}^T:\ell_R^T \ne 0} \mathbb{E}\left[\Delta_{\pmb{d}^T}^2(\Delta_{max})\right] \pi_0'\left(\pmb{0},\pmb{\ell}^T\right), \notag \\
\mathbb{E}\left[\Delta_{\pmb{d}^T}^1(\Delta_{max})\right] &= \sum_{\Delta^1=1}^{\Delta_{max}} f_{\pmb{D}^T}\left(\pmb{d}^T|\Delta^1\right), \notag \\
\mathbb{E}\left[\Delta_{\pmb{d}^T}^2(\Delta_{max})\right] &= \sum_{\Delta^1=1}^{\Delta_{max}} \sum_{\Delta^2=\Delta^1}^{\Delta_{max}} f_{\pmb{D}^T}\left(\pmb{d}^T|\Delta^1,\Delta^2\right), \notag
\end{align}
\noindent where $f_{\pmb{D}^T}\left(\pmb{d}^T|\Delta^1\right)$ is the terminal state distribution in phase $1$, which has initial state distribution $f_{\pmb{D}^{I_1}}\left(\pmb{d}^{I_1}|\Delta^1\right)$, and $f_{\pmb{D}^T}\left(\pmb{d}^T|\Delta^1,\Delta^2\right)$ is the terminal state distribution in phase $2$, which has initial state distribution $f_{\pmb{D}^{I_2}}\left(\pmb{d}^{I_2}|\Delta^1,\Delta^2\right)$.\qed
\end{theorem}

The initial state distribution for the first phase, $f_{\pmb{D}^{I_1}}\left(\pmb{d}^{I_1}|\Delta^1\right)$, is found using \eqref{init_state}. For the second phase, we note that the initial state distribution, $f_{\pmb{D}^{I_2}}\left(\pmb{d}^{I_2}|\Delta^1,\Delta^2\right)$, will be the result of receiving an additional $\Delta^2-\Delta^1$ symbols and identifying the initial ripple, while being in a state where the feedback was transmitted, which is referred to as a \textit{feedback state} and denoted $\pmb{D}^F=[\pmb{L}^F\text{ }\pmb{R}^F\text{ }\pmb{C}^F]$. The outcome leading to an initial state for phase $2$ can be viewed as the combination of four events, $E_1$, $E_2$, $E_3$ and $E_4$, which are defined below:

\begin{align}
E_1: & \quad{} \text{The terminal state of the first phase is a feedback state, $\pmb{D}^F$.} \notag \\
E_2: & \quad{} \text{The new $\Delta^2-\Delta^1$ symbols have original degrees as expressed by $\pmb{\omega^2}$, where $\omega^2_i$ is the} \notag \\
     & \quad{} \text{number of new symbols of original degree $i$.} \notag \\
E_3: & \quad{} \text{For each $i$, out of the $\omega^2_i$ new symbols of original degree $i$, $m_{Ri}^{I_2}$ have reduced degree $1$,} \notag \\
     & \quad{} \text{and $c^{F}_i+\omega^2_i-m_{Ri}^{I_2}-c_i^{I_2}$ have reduced degree $0$.} \notag \\
E_4: & \quad{} \text{Out of the $\hat{m}_{R}^{I_2}$ released symbols, $r^{I_2}_R$ are added to the ripple.} \notag
\end{align}

These events are not independent, since $E_3$ depends on $E_1$ and $E_2$, and $E_4$ depends on $E_1$, $E_2$ and $E_3$. We can express the initial state distribution for phase $2$ as follows:

\begin{align}
f_{\pmb{D}^{I_2}}\left(\pmb{d}^{I_2}|\Delta^1,\Delta^2\right)  &= \sum_{\pmb{d}^{F} \in {\cal{S}}_1} \sum_{\pmb{\omega^2} \in {\cal{S}}_2} \sum_{\pmb{m}_R^{I_2} \in {\cal{S}}_3} \text{Pr}(E_1)\text{Pr}(E_2)\text{Pr}(E_3|E_1,E_2)\text{Pr}(E_4|E_1,E_2,E_3) \notag \\
                                             {\cal{S}}_1 &\triangleq \{ \pmb{d}^{F} : c^{F}_i\le c^{I_2}_i,\text{ }\forall i\}, \notag \\
                                             {\cal{S}}_2 &\triangleq \{ \pmb{\omega^2} : \hat{\omega}^2=\Delta^2-\Delta^1 \text{ and } c^{F}_i+\omega^2_i \ge c^{I_2}_i,\text{ }\forall i\}, \notag \\
                                             {\cal{S}}_3 &\triangleq \{ \pmb{m}_R^{I_2} : \hat{m}_{R}^{I_2}\ge r^{I_2}_R \text{ and } m_{Ri}^{I_2} \le \omega^2_i,\text{ }\forall i\},
\end{align}

where the sets ${\cal{S}}_1$, ${\cal{S}}_2$ and ${\cal{S}}_3$ refers to all possible events $E_1$, $E_2$ and $E_3$, which enable an event $E_4$ that provides $\pmb{d}^{I_2}$.

The probability of $E_1$ is found by evaluating the feedback state distribution, $f_{\pmb{D}^{F}}\left(\pmb{d}^{F}|\Delta^1\right)$. A feedback state is defined as a terminal state in which $L_B=0$, provided that the decoding attempt of $\Delta^1-1$ symbols resulted in a terminal state in which $L_B>0$. Hence, the $\Delta$'th received symbol must have reduced degree $1$ and enable the decoder to recover the remaining base layer symbols. By $f_{\pmb{D}^{F-}}\left(\pmb{d}^{F-}|\Delta^1\right)$, we denote the state distribution after the decoding of $\Delta^1-1$ symbols and receiving the $\Delta$'th symbol, which is potentially added to the ripple. We can then express the feedback state distribution, and thereby the probability of $E_1$, as follows:

\begin{align}
\text{Pr}(E_1) &= f_{\pmb{D}^{F}}\left(\pmb{d}^{F}|\Delta^1\right) = \left\{ \begin{array}{ll} f_{\pmb{D}^{T}}\left(\pmb{d}^{F}|\Delta^1\right), &\text{for }\ell^F_B=0, \\ 0, &\text{elsewhere}, \end{array} \right. \notag \\
\end{align}

\noindent where the initial state distribution of the state recursion used to express $f_{\pmb{D}^{T}}\left(\pmb{d}^{F}|\Delta^1\right)$ is given by $f_{\pmb{D}^{F-}}\left(\pmb{d}^{F-}|\Delta^1\right)$, which is found as follows:

\begin{align}
f_{\pmb{D}^{F-}}\left(\pmb{d}^{F-}|\Delta^1\right)    &= \left\{ \begin{array}{ll} f_{\pmb{D}^{T}}\left([ \pmb{\ell}^{F-}\text{ }\pmb{0}\text{ }\pmb{c}^{F-}] |\Delta^1-1\right) \pi_{\beta}'\left(\pmb{r}^{F-}|\pmb{\ell}^{F-}\right), &\text{for }\ell^{F-}_B>0,\text{ }\hat{r}^{F-}=1, \\ 0, &\text{elsewhere}. \end{array} \right.
\end{align}

The probability of $E_2$ is found using the multinomial distribution, since the original degrees are sampled from $\pi(i)$. Hence,

\begin{align}
\text{Pr}(E_2)= \mu\left(\pmb{\omega^2},\Delta^2-\Delta^1,\pi(i)\right).
\end{align}

$E_3$ concerns the reduced degree, $i_R'$, which follows the hypergeometric distribution, $\Upsilon(i_R',(1-\alpha)k,\ell_R^F,i)$, since new output symbols are encoded using only refinement layer symbols. The probability that a symbol with original degree $i$ has reduced degree $0$ $(1)$ is denoted $p_0(i)$ $(p_1(i))$. Then for any $i$, $m_{Ri}^{I_2}$ follows the binomial distribution, $\theta\left(m_{Ri}^{I_2},\omega^2_i,p_1(i)\right)$. The remaining $\omega^2_i-m_{Ri}^{I_2}$ symbols have reduced degree $0$ with conditional probability $\frac{p_0(i)}{1-p_1(i)}$. For a given $c_i^{I_2}$ and $m_{Ri}^{I_2}$, we have that the number of symbols of reduced degree $0$ must be $c^{F}_i+\omega^2_i-m_{Ri}^{I_2}-c_i^{I_2}$, since the total amount of symbols with original degree $i$ is $c^{F}_i+\omega^2_i$. The probability of this is $\theta\left(c^{F}_i+\omega^2_i-m_{Ri}^{I_2}-c_i^{I_2},\omega^2_i-m_{Ri}^{I_2},\frac{p_0(i)}{1-p_1(i)}\right)$. The probability of $E_3$ is then found as a multiplication of the two binomials for all $i$. Hence,

\begin{align}
\text{Pr}(E_3|E_1,E_2)= \prod_{i=1}^k \bigg( \theta\left(m_{Ri}^{I_2},\omega^2_i,p_1(i)\right) \theta\left(c^{F}_i+\omega^2_i-m_{Ri}^{I_2}-c_i^{I_2},\omega^2_i-m_{Ri}^{I_2},\frac{p_0(i)}{1-p_1(i)}\right) \bigg).
\end{align}

For a given $E_3$, the vector $\pmb{m}_R^{I_2}$ is given and thereby $\hat{m}_{R}^{I_2}$, the total amount of released symbols. The probability that $r^{I_2}_R$ of these are added to the ripple is found using Lemma \ref{ripadds}. Hence,

\begin{align}
\text{Pr}(E_4|E_1,E_2,E_3)= f_{\hat{M}'_n}\left(r^{I_2}_R|\hat{m}_{R}^{I_2},\ell^F_R,0\right).
\end{align}

This concludes the analysis and we are now able to evaluate the presented theorems. This is described in the following section.

\section{Numerical Results} \label{sec:results}
In this section we present evaluations of the expressions derived in section \ref{sec:analysis} as well as simulations of a URT-LT code. All evaluations are performed with $k=100$, $\alpha=0.5$ and the RSD as the degree distribution, with parameters $c=0.1$ and $\delta=1$. To evaluate the expressions, we use Monte Carlo simulations with $1000$ iterations. 

\begin{figure}[t]
\centering
\includegraphics[width=\columnwidth]{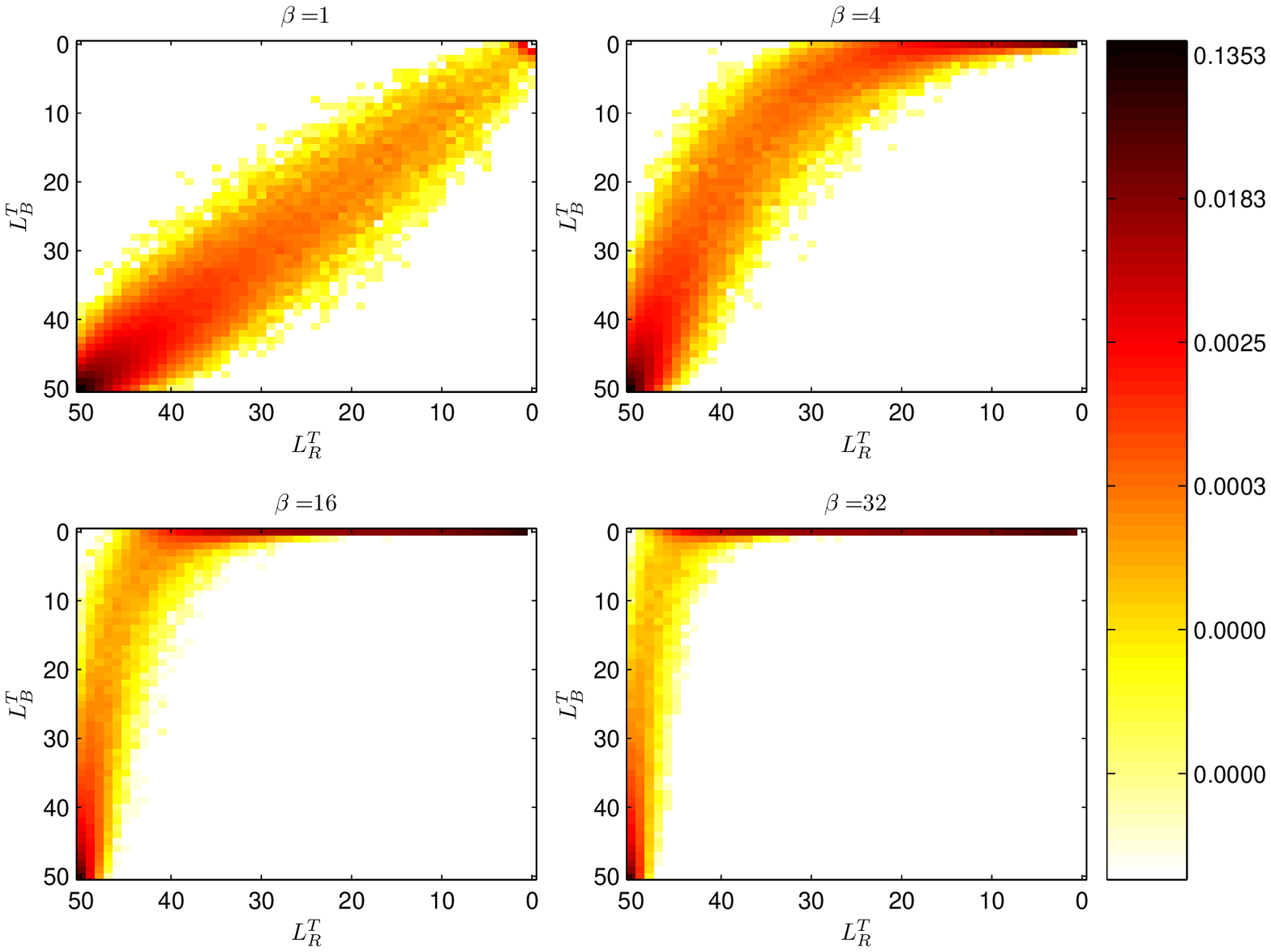}
\caption{Normalized $E\left[\Delta_{\pmb{d}^T}\right]$ as a function of $L_B^T$ and $L_R^T$ at different values of $\beta$.}
\label{fig:betas_nack}
\end{figure}

\begin{figure}[t]
\centering
\includegraphics[width=\columnwidth]{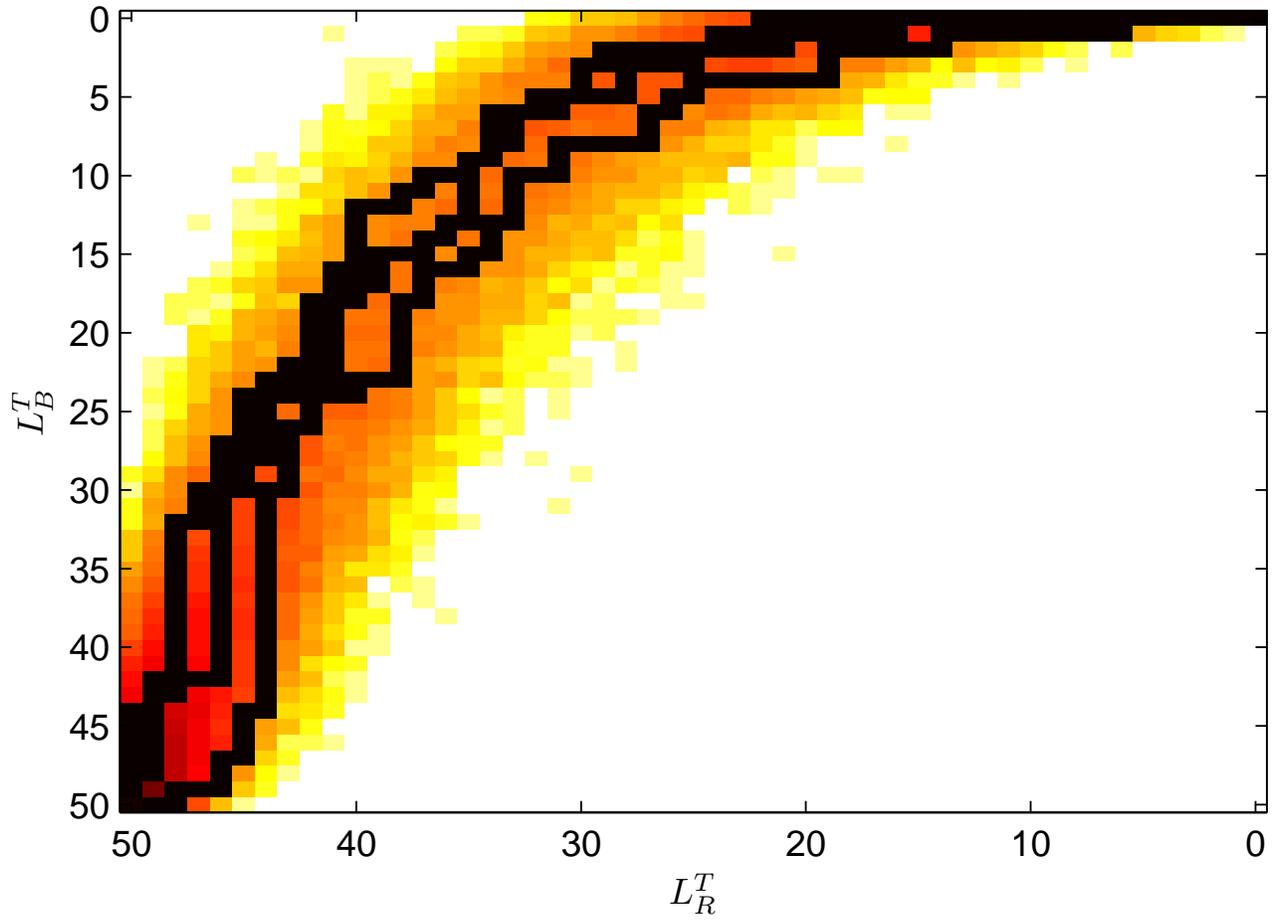}
\caption{Actual simulations of the analyzed URT-LT code illustrated on top of the analytical results.}
\label{fig:runs}
\end{figure}

\begin{figure}[t]
\centering
\includegraphics[width=\columnwidth]{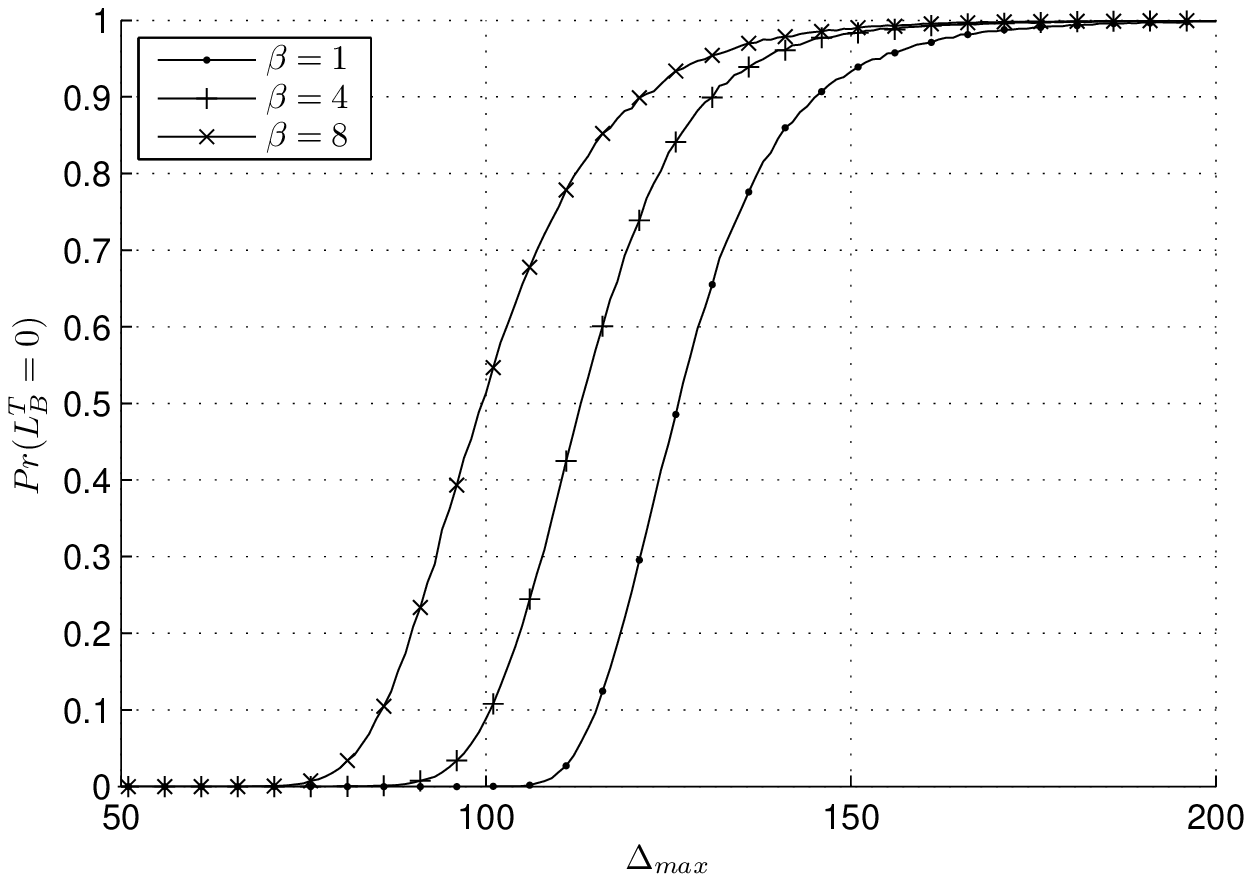}
\caption{Probability of successfully decoding the base layer as a function of the maximum amount of collected symbols.}
\label{fig:urt}
\end{figure}

Initially, we evaluate $E\left[\Delta_{\pmb{d}^T}(\Delta_{max})\right]$, i.e. the expected amount of symbols received in a given terminal state, $\pmb{d}^T$, at different values of $\beta$ and in the case of no feedback. We evaluate it at all possible $\pmb{D}^T$ and normalize with the maximum number of received symbols, $\Delta_{max}$, hence providing the expected fraction of symbols received in that given state. Moreover we marginalize out $\pmb{R}^T$ and $\pmb{C}^T$, which means we get the expected fraction as a function of $\pmb{L}^T$. The results are illustrated in Fig. \ref{fig:betas_nack} for $\beta=\{1,4,16,32\}$. Note that the color code is using a logarithmic scale, in order to better visualize the results. From the figure it is seen that at $\beta=1$, symbol receptions are distributed symmetrically around the line $L_B^T=L_R^T$, which was expected, since at $\beta=1$ we have a standard LT code with no bias towards the base layer. In this case, we also see that most symbols are received in states where none or very few input symbols have been recovered. This confirms the well-known avalanche effect in LT decoding \cite{mackay}, which refers to the fact that the first many received symbols only enable the recovery of very few input symbols. Then suddenly, a single new symbol enables the recovery of all the remaining input symbols. A brief look at Fig. \ref{fig:twolayerred} reveals that this effect is essential to the performance of standard LT codes. Moving on to higher $\beta$ values, we see the bias towards the base layer come into effect. Clearly, symbols are more likely to be received in states where $L_B^T<L_R^T$, which is an indication of the URT property. However, it is also clear that this bias results in more symbols being received in states where $L_B^T=0$ and $L_R^T>0$. In other words, the avalanche fades out prematurely and new symbols must be received in a state where few symbols are unrecovered, leading to high probability of redundancy, cf. Fig. \ref{fig:twolayerred}. The case of $\beta=4$ with three simulations, not analytical results, of the corresponding URT-LT code added as an extra layer on top is shown in Fig. \ref{fig:runs}. In the simulations $L_B$ and $L_R$ have been logged during a successful decoding. This figure illustrates that the behavior seen in Fig. \ref{fig:betas_nack} corresponds well with practice. The fact that the URT property is indeed achieved with these codes is verified by Fig. \ref{fig:urt}, where the probability of having successfully decoded the base layer is plotted as a function of $\Delta$. Note that this probability is found by evaluating $f_{\pmb{D}^{T}}\left(\pmb{d}^{T},\Delta\right)$ at $L_B^T=0$ and marginalizing out all other dimensions of the state space.

\begin{figure}[t]
\centering
\includegraphics[width=\columnwidth]{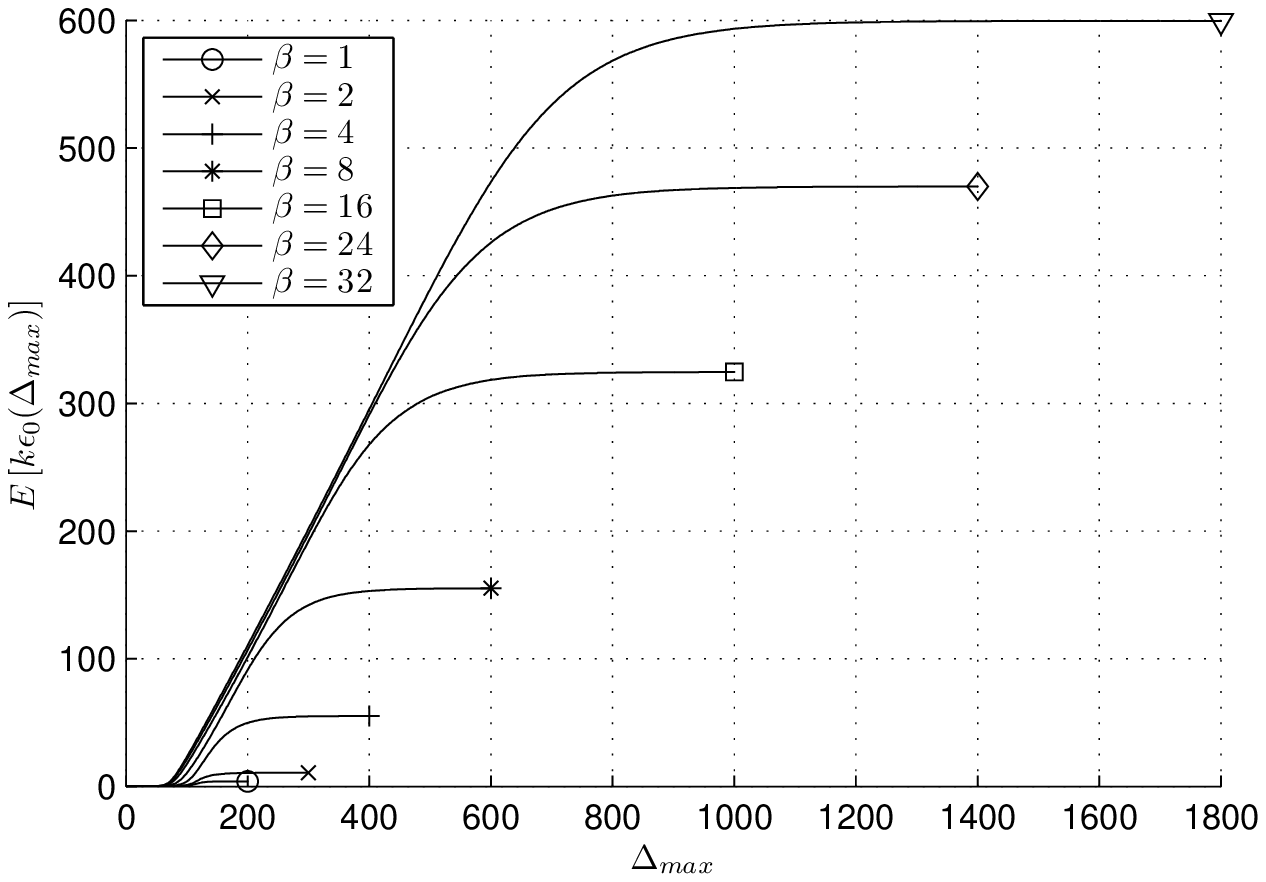}
\caption{Expected amount of redundancy due to reduced degree zero as a function of the maximum amount of collected symbols for different values of $\beta$ in the case of no feedback.}
\label{fig:nack}
\end{figure}

\begin{figure}[t]
\centering
\includegraphics[width=\columnwidth]{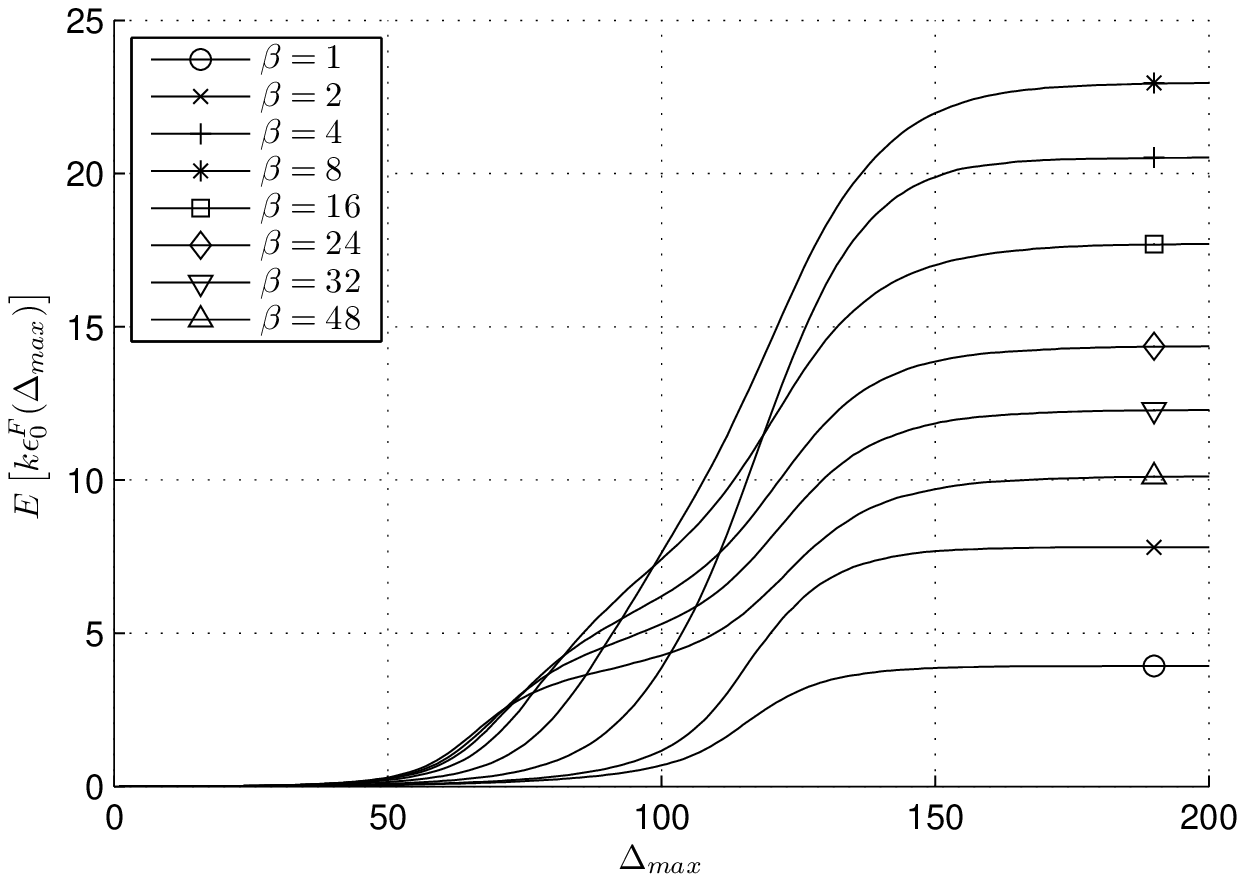}
\caption{Expected amount of redundancy due to reduced degree zero as a function of the maximum amount of collected symbols for different values of $\beta$ in the case of feedback.}
\label{fig:wack}
\end{figure}

\begin{figure}[t]
\centering
\includegraphics[width=\columnwidth]{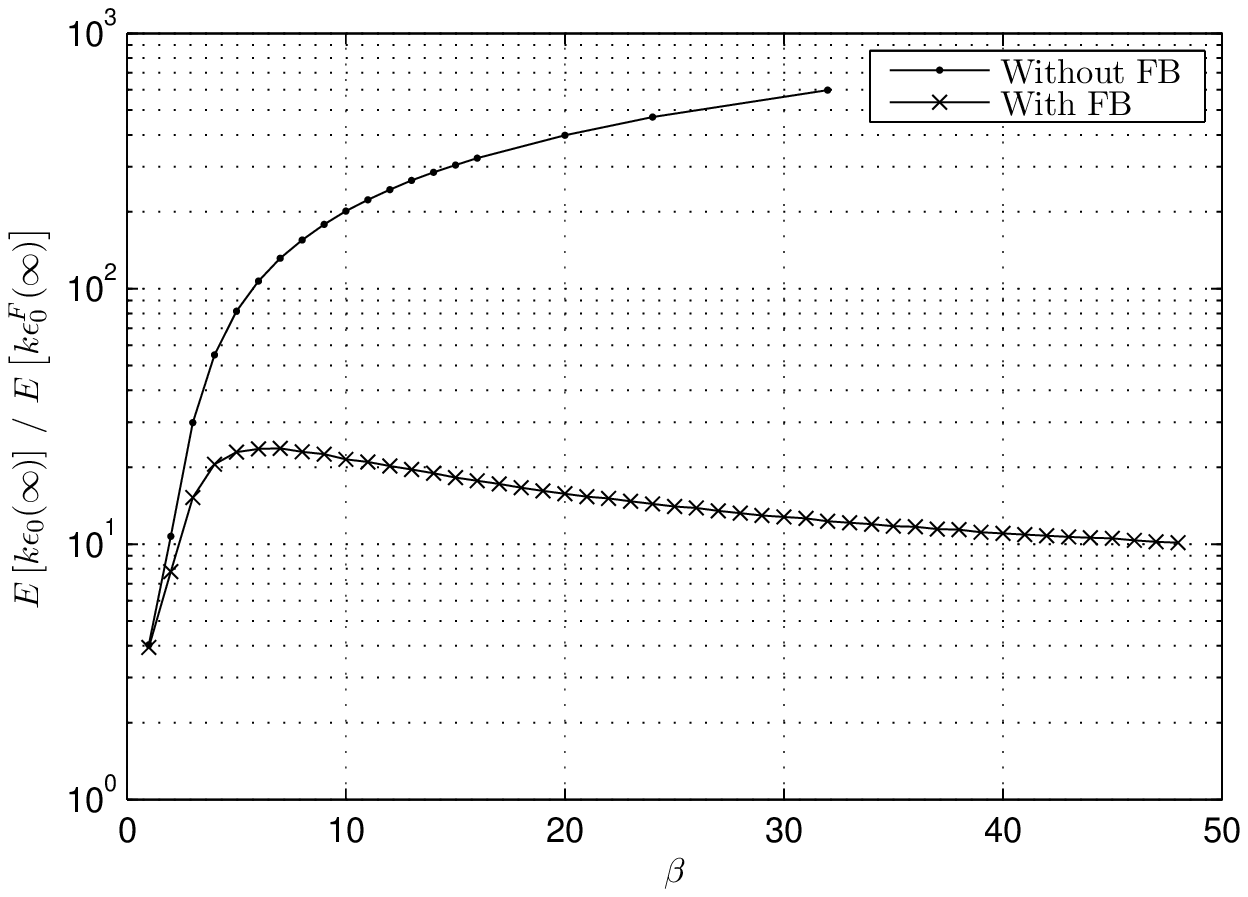}
\caption{Asymptotic values of $\mathbb{E}\left[k \epsilon_0(\Delta_{max})\right]$ and $\mathbb{E}\left[k \epsilon_0^F(\Delta_{max}) \right]$ as a function of $\beta$.}
\label{fig:asymp}
\end{figure}

Next, we evaluate $\mathbb{E}\left[k \epsilon_0(\Delta_{max})\right]$ and $\mathbb{E}\left[k \epsilon_0^F(\Delta_{max}) \right]$, again for increasing values of $\beta$. The results are shown in Fig. \ref{fig:nack} (no feedback) and Fig. \ref{fig:wack} (with feedback) as a function of $\Delta_{max}$. In general in both figures, we see that the amount of redundancy remains close to zero until roughly $50$ symbols have been received, after which redundancy starts to occur. The amount of redundancy increases until all input symbols have been recovered ($\pmb{L}^T=\pmb{0}$) with high probability and converges to a level, that depends upon $\beta$. In the case of no feedback, it is seen that the redundancy increases linearly with $\beta$. Hence, the URT property comes at a significant price in the form of additional overhead and this price increases with the bias towards the base layer. In the case where feedback is used to acknowledge the base layer, we also see an increase in redundancy for increasing $\beta$. However, the redundancy reaches a maximum at $\beta=8$ and then starts to decrease for $\beta>8$. This clearly demonstrates the great potential a single intermediate feedback has in URT-LT codes. Fig. \ref{fig:asymp} shows the converged redundancy values, i.e. $\mathbb{E}\left[k \epsilon_0(\infty)\right]$ and $\mathbb{E}\left[k \epsilon_0^F(\infty) \right]$, as a function of $\beta$. This makes it easy to compare the two schemes. Note that the redundancy converges for increasing $\beta$ in the scheme applying intermediate feedback. This is due to the fact that this scheme, in the limit $\beta=\infty$, is the equivalent of two standard LT coded transmissions, with $k_1=\alpha k$ and $k_2=(1-\alpha)k$.


\begin{figure}[t]
\centering
\includegraphics[width=\columnwidth]{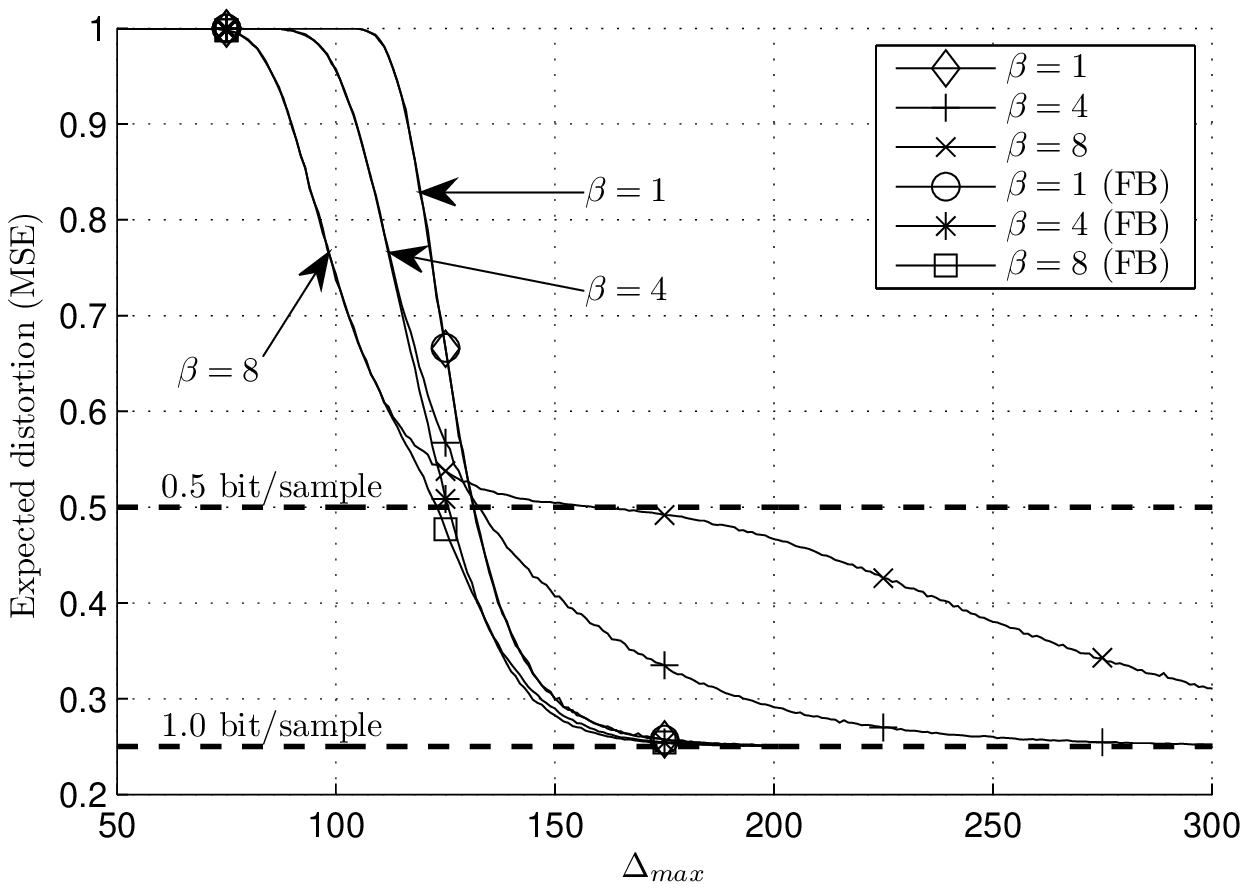}
\caption{Expected distortion of a layered unit-variance Gaussian source as a function of the maximum amount of collected symbols.}
\label{fig:rd}
\end{figure}

Finally, we present an evaluation where we map terminal states to distortion measures. We assume that the decoded data describes a successive refinable \cite{cover} Gaussian source with unit variance, which has been partitioned into two layers; base layer and refinement layer. It is assumed that if both layers are decoded ($L_B^T=L_R^T=0$), $1$ bit/sample is available to describe the source. If only the base layer is decoded ($L_B^T=0,L_R^T>0$), $\alpha$ bit/sample is available. If either no layers ($L_B^T=L_R^T>0$) or the refinement layer only ($L_B^T>0,L_R^T=0$) is decoded, we have $0$ bit/sample to describe the source. This layered structure, which is common in e.g. video streaming, motivates the use of URT-LT codes, since the base layer has more importance with respect to the distortion measure. For the unit variance Gaussian source, we have that $D \ge 2^{(-2R)}$, where $D$ is the distortion measured as the mean squared error and $R$ is the rate measured in bits per sample. We assume that the bound is achievable and use this relationship to calculate the expected distortion of the URT-LT codes for both the case with acknowledgment of the base layer and the case without. The results are shown in Fig. \ref{fig:rd} for the same code parameters as in previous evaluations. The figure shows that the optimal value of $\beta$ depends on the overhead of the URT-LT code. Moreover, it is evident that in the case of no acknowledgment of the base layer, if a low distortion is required at low overhead, the price to pay is a very significant increase of the distortion at higher overhead. If the base layer is acknowledged, the distortion quickly decreases to the minimum level, regardless of the choice of $\beta$.

\section{Conclusions}\label{sec:conclusion}
We have analyzed finite-length LT codes with unequal recovery time, termed URT-LT codes in this paper. The analysis is based on a state recursion function, which allows us to evaluate the distributions of the ripple, cloud and decoding progress in individual segments, as the overall decoding progresses. This gives novel insight into the probabilistic mechanisms in URT-LT codes, which is a major contribution of this paper. The analysis enables us to evaluate the expected amount of symbols with reduced degree $0$, i.e. redundant symbols, during a transmission. Evaluations in the case of two data segments show that this amount increases roughly linearly with the level of priority given to the first data segment. As a result, successful decoding of the lower prioritized data is delayed substantially. Thus, we can conclude that the unequal recovery time comes at a significant price in terms of redundancy in lower prioritized data. 

A slight modification of the URT-LT codes has been proposed, where an intermediate feedback message informs the encoder that the higher prioritized data has been decoded. The encoder adapts by excluding the decoded data from future encoding. Analysis of this code reveals that such a modification is able to dramatically decrease the redundancy in lower prioritized data. The impact of this improvement is further illustrated with an evaluation, where the decoding probabilities are mapped to an expected distortion measure of a layered unit variance Gaussian source. Improvements of roughly $0.25$ in mean squared error is observed compared to the original URT-LT code.
\appendix[]
\begin{IEEEproof}[Proof of Theorem \ref{nrdd}]
Initially it is noted that $\pmb{j}$ follows Wallenius' multivariate noncentral hypergeometric distribution with parameters $i$, $\pmb{\alpha}k$ and $\pmb{\beta}$. This distribution generalizes the hypergeometric distribution to take nonuniform sampling into account. For a certain $\pmb{j}$, the probability of receiving a symbol with reduced degree $\pmb{i'}$ is found as a product of $N$ regular hypergeometric distributions. This follows from the fact that a degree reduction from $j_n$ to $i_n'$, $n=1,2,...,N$, occurs when $i_n'$ of the neighbors are among the $\ell_n$ undecoded symbols and the remaining $j_n-i_n'$ neighbors are among the $\alpha_n k-\ell_n$ already decoded symbols. Sampling of neighbors within a single layer is done uniformly, hence the hypergeometric distribution applies. Finally it is noted that any symbol with $\pmb{j}\in\mathcal{J}_i$ can potentially reduce to $\pmb{i'}$ since $j_n \ge i'_n$, $n=1,2,...,N$, and that $\mathcal{J}_i=\emptyset$ unless $\hat{i}'\le i \le \hat{i}'+k-\hat{L}$. 
\end{IEEEproof}

\begin{IEEEproof}[Proof of Theorem \ref{red1}]
The probability of ending in the terminal state $\pmb{d}^T$, thus receiving the next symbol in this state, when trying to decode the first $\Delta$ symbols, is given by $f_{\pmb{D}^T}\left(\pmb{d}^T,\Delta\right)$. The expected number of symbols, $\mathbb{E}\left[\Delta_{\pmb{d}^T}(\Delta_{max})\right]$, received in this state during an entire transmission is found by summing $f_{\pmb{D}^T}\left(\pmb{d}^T,\Delta\right)$ for all possible $\Delta$, i.e. $\Delta=1,...,\Delta_{max}$. Of these, an expected $\mathbb{E}\left[\Delta_{\pmb{d}^T}(\Delta_{max})\right] \pi_{\beta}'\left(\pmb{0},\pmb{\ell}^T\right)$ have reduced degree $0$. Finally, the total amount of symbols, $\mathbb{E}\left[\epsilon_0 k\right]$, with reduced degree $0$ is found by summing over all terminal states in which the transmission continues, i.e. any state in which $\pmb{\ell}^T\neq\pmb{0}$.
\end{IEEEproof}

\begin{IEEEproof}[Proof of Theorem \ref{red2}]
The contribution from phase $1$, $\sum_{\pmb{d}^T:\ell_B^T \ne 0} \mathbb{E}\left[\Delta_{\pmb{d}^T}^1(\infty)\right] \pi_{\beta}'\left(\pmb{0},\pmb{\ell}^T\right)$, follows the same structure as Theorem \ref{red1}, although with a different condition for receiving more symbols, since phase $1$ continues as long as $\ell_B^T \ne 0$. The same holds for the contribution from phase $2$, $\sum_{\pmb{d}^T:\ell_R^T \ne 0} \mathbb{E}\left[\Delta_{\pmb{d}^T}^2(\infty)\right] \pi_0'\left(\pmb{0},\pmb{\ell}^T\right)$, although the amount of symbols received in phase $2$ is $\Delta_2-\Delta_1$ and the condition for receiving more symbols is $\ell_R^T \ne 0$. See proof of Theorem \ref{red1} for details.
\end{IEEEproof}

\begin{IEEEproof}[Proof of Lemma \ref{ripadds}]
If $\hat{m}_n$ samples are drawn uniformly at random from a set of size $\ell_n$, then the number of unique samples, $q$, will follow a distribution expressed by $\frac{\binom{\ell_n}{q} Z_q(\hat{m}_n)}{(\ell_n)^{\hat{m}_n}}$, where $Z_q(\hat{m}_n) = \sum_{p=0}^q \binom{q}{p} (q-p)^{\hat{m}_n}(-1)^p$. This is an inverse variant of the coupon collector's problem, whose details can be found in \cite{}. Having $q$ unique released symbols, the number of those, $\hat{m}_n'$, who are among the $\ell_n-r_n$ symbols not already in the ripple follows the hypergeometric distribution. Having $\hat{m}_n'$ ripple additions can be the result of any amount of unique releases higher than $\hat{m}_n'$.
\end{IEEEproof}

\begin{IEEEproof}[Proof of Lemma \ref{nextproc}]
Since the next symbol to be processed is chosen uniformly at random among the symbols in the ripple, the probability that it is a base layer symbol is the fraction of base layer symbols currently in the ripple, $\frac{r^{\hat{L}+1}_B}{r^{\hat{L}+1}_B+r^{\hat{L}+1}_R}$. Similarly, the probability that it is a refinement layer symbol is $\frac{r^{\hat{L}+1}_R}{r^{\hat{L}+1}_B+r^{\hat{L}+1}_R}$.
\end{IEEEproof}

\begin{IEEEproof}[Proof of Lemma \ref{relprob}]
Given a degree $i$, $\Phi(j,i,\alpha k,\beta)$ expresses the probability of having $j$ base layer symbols among the $i$ neighbors. See proof of Theorem \ref{nrdd} for details. Assuming a base layer symbol has just been processed, a symbol is released as a new base layer symbol when $\ell_B$ and $\ell_R$ symbols remain unprocessed from base layer and refinement layer respectively, if $j-2$ base layer neighbors are among the $\alpha k-\ell_B-1$ first processed base layer symbols, one is the $(\alpha k-\ell_B)$'th processed base layer symbol, all $i-j$ refinement layer neighbors are among the $(1-\alpha) k-\ell_R$ processed refinement layer symbols and the last base layer neighbor is among the $\ell_B$ undecoded base layer symbols. This proves the expression for $q_{BB}(i,\pmb{\ell}^{\hat{L}})$. Similar proofs can be made for $q_{BR}(i,\pmb{\ell}^{\hat{L}})$, $q_{RB}(i,\pmb{\ell}^{\hat{L}})$ and $q_{RR}(i,\pmb{\ell}^{\hat{L}})$.
\end{IEEEproof}

\begin{IEEEproof}[Proof of Lemma \ref{rels}]
Any symbol with original degree $i$ is released in the $(k-\hat{L})$'th decoding step with prior probability $q\left(i,\pmb{\ell}^{\hat{L}}\right)$ according to Lemma \ref{relprob}. All symbols with original degree $i$ still left in the cloud after $k-(\hat{L}+1)$ decoding steps will release with conditional probability $q_c\left(i,\pmb{\ell}^{\hat{L}}\right) = \frac{q\left(i,\pmb{\ell}^{\hat{L}}\right)}{\sum_{\ell_R=0}^{\ell_R^{\hat{L}}} q\left(i,[0\text{ }\ell_R]\right) + \sum_{\ell_B=1}^{\ell_B^{\hat{L}}} q\left(i,[\ell_B\text{ }\ell_R^{\hat{L}}]\right)}$, where the denominator expresses the remaining probability mass of the prior probability distribution from Lemma \ref{relprob}. Hence, the amount of degree $i$ symbols, $\hat{M}^{\hat{L}}_i$, released in the next decoding step, follows the binomial distribution, $\theta\left(\hat{m}^{\hat{L}}_i,c^{\hat{L}+1}_i,q_c\left(i,\pmb{\ell}^{\hat{L}}\right)\right)$. The probability of having a cloud of $\pmb{c}^{\hat{L}}$ after that decoding step is thus found as a product of binomial distributions, evaluated at $\hat{m}^{\hat{L}}_i$, $\forall$ $i$.
\end{IEEEproof}

\begin{IEEEproof}[Proof of Lemma \ref{ripdev}]
For a certain cloud development in the $(k-\hat{L})$'th decoding step, the total amount of releases, $\hat{m}^{\hat{L}}$, is found as $\hat{m}^{\hat{L}} = \sum_{i=2}^k c^{\hat{L}+1}_i-c^{\hat{L}}_i$. These releases are differentiated among layers using equation \eqref{mb}, thereby achieving $\hat{m}^{\hat{L}}_{B}$ and $\hat{m}^{\hat{L}}_{B}=\hat{m}^{\hat{L}}-\hat{m}^{\hat{L}}_{B}$. Lemma \ref{ripadds} is then used to express the distributions of the amounts, $\hat{M}^{\hat{L}'}_{B}$ and $\hat{M}^{\hat{L}'}_{R}$, which are added to the ripple.


\end{IEEEproof}



\bibliographystyle{ieeetr}
\bibliography{bibliography}

\begin{thebibliography}{10}

\bibitem{fc2}
{M. Luby}, ``{LT Codes},'' in {\em {Proceedings. The 43rd Annual IEEE Symposium
  on Foundations of Computer Science.}}, pp.~{271--280}, {November} 2002.

\bibitem{raptor}
{A. Shokrollahi}, ``{Raptor codes},'' {\em {IEEE Transactions on Information
  Theory.}}, pp.~{2551--2567}, 2006.

\bibitem{blostein}
Y.~Cao, S.~Blostein, and W.-Y. Chan, ``Optimization of rateless coding for
  multimedia multicasting,'' in {\em Broadband Multimedia Systems and
  Broadcasting (BMSB), 2010 IEEE International Symposium on}, pp.~1--6, March
  2010.

\bibitem{nybom}
K.~Nybom, S.~Gr\"{o}nroos, and J.~Bj\"{o}rkqvist, ``Expanding window fountain
  coded scalable video in broadcasting,'' in {\em Multimedia and Expo (ICME),
  2010 IEEE International Conference on}, pp.~516--521, July 2010.

\bibitem{karande}
S.~Karande, K.~Misra, S.~Soltani, and H.~Radha, ``Design and analysis of
  generalized lt-codes using colored ripples,'' in {\em Information Theory,
  2008. ISIT 2008. IEEE International Symposium on}, pp.~2071--2075, July 2008.

\bibitem{bogino}
M.~Bogino, P.~Cataldi, M.~Grangetto, E.~Magli, and G.~Olmo, ``Sliding-window
  digital fountain codes for streaming of multimedia contents,'' in {\em
  Circuits and Systems, 2007. ISCAS 2007. IEEE International Symposium on},
  pp.~3467--3470, May 2007.

\bibitem{dejan}
D.~Sejdinovi\'{c}, D.~Vukobratovi\'{c}, A.~Doufexi, V.~\v{S}enk, and
  R.~Piechocki, ``Expanding window fountain codes for unequal error
  protection,'' {\em Communications, IEEE Transactions on}, vol.~57, pp.~2510
  --2516, september 2009.

\bibitem{neto}
H.~Neto, W.~Henkel, and V.~da~Rocha, ``Multi-edge framework for unequal error
  protecting lt codes,'' in {\em Information Theory Workshop (ITW), 2011 IEEE},
  pp.~267--271, October 2011.

\bibitem{rahnavard}
N.~Rahnavard, B.~Vellambi, and F.~Fekri, ``Rateless codes with unequal error
  protection property,'' {\em Information Theory, IEEE Transactions on},
  vol.~53, pp.~1521--1532, April 2007.

\bibitem{cataldi}
P.~Cataldi, M.~Grangetto, T.~Tillo, E.~Magli, and G.~Olmo, ``Sliding-window
  raptor codes for efficient scalable wireless video broadcasting with unequal
  loss protection,'' {\em Image Processing, IEEE Transactions on}, vol.~19,
  pp.~1491--1503, June 2010.

\bibitem{karp}
R.~Karp, M.~Luby, and A.~Shokrollahi, ``Finite length analysis of lt codes,''
  in {\em Information Theory, 2004. ISIT 2004. Proceedings. International
  Symposium on}, p.~39, June/July 2004.

\bibitem{uepfc1}
{N. Rahnavard, B. N. Vellambi and F. Fekri}, ``{Rateless Codes With Unequal
  Error Protection Property},'' {\em {IEEE Transactions on Information Theory
  vol. 53.}}, pp.~{1521 -- 1532}, {April} 2007.

\bibitem{mackay}
D.~MacKay, ``Fountain codes,'' {\em Communications, IEE Proceedings-},
  vol.~152, pp.~1062 -- 1068, December 2005.

\bibitem{cover}
T.~Cover and J.~Thomas, {\em Elements of information theory}.
\newblock New York: Wiley, 1991.

\end{thebibliography}

\end{document}